\documentclass[useAMS,usenatbib]{mn2e}
\newcommand{\bib}{\bibitem[\protect\citeauthoryear}
\usepackage[dvips]{graphicx}
\usepackage[T1]{fontenc}
\usepackage{ae,aecompl}

\title[NGC\,315 jet model]{A relativistic model of the radio jets in NGC\,315}
\author[J.R. Canvin et al.]{J.R. Canvin \thanks{E-mail:
jcanvin@physics.usyd.edu.au}$^{1,2}$ , R.A. Laing$^{3}$, A.H.Bridle$^{4}$,
W.D. Cotton$^{4}$\\$^1$ School of Physics, University of Sydney, A28, Sydney,
NSW 2006, Australia\\ $^2$ University of Oxford, Department of Astrophysics,
Denys Wilkinson Building, Keble Road, Oxford OX1 3RH \\ $^3$ European Southern
Observatory, Karl-Schwarzschild-Stra\ss e 2, D-85748 Garching-bei-M\"{u}nchen,
Germany\\ $^4$ National Radio Astronomy Observatory, 520 Edgemont Road,
Charlottesville, VA 22903-2475, U.S.A.}

\date{Received}

\pagerange{\pageref{firstpage}--\pageref{lastpage}} \pubyear{2005}

\begin{document}
\label{firstpage}

\maketitle

%-------------------------- ****** ABSTRACT ****** --------------------------%
\begin{abstract}
We apply our intrinsically symmetrical, decelerating relativistic jet model to
deep VLA imaging of the inner $\pm$70\,arcsec of the giant low-luminosity radio
galaxy NGC\,315. An optimized model accurately fits the data in both total
intensity and linear polarization. We infer that the velocity, emissivity and
field structure in NGC\,315 are very similar to those of the other
low-luminosity sources we have modelled, but that all of the physical scales are
larger by a factor of about 5.  We derive an inclination to the line of sight of
$38^\circ \pm 2^\circ$ for the jets. Where they first brighten, their on-axis
velocity is $\beta = v/c \approx 0.9$. They decelerate to $\beta \approx 0.4$
between 8 and 18\,kpc from the nucleus and the velocity thereafter remains
constant.  The speed at the edge of the jet is $\approx$0.6 of the on-axis value
where it is best constrained, but the transverse velocity profile may deviate
systematically from the Gaussian form we assume. The proper emissivity profile
is split into three power-law regions separated by shorter transition zones. In
the first of these, at $\approx$3\,kpc (the flaring point) the jets expand
rapidly at constant emissivity, leading to a large increase in the observed
brightness on the approaching side. At $\approx$10\,kpc, the emissivity drops
abruptly by a factor of 2. Where the jets are well resolved their rest-frame
emission is centre-brightened.  The magnetic field is modelled as random on
small scales but anisotropic and we rule out a globally ordered helical
configuration. To a first approximation, the field evolves from a mixture of
longitudinal and toroidal components to predominantly toroidal, but it also
shows variations in structure along and across the jets, with a significant
radial component in places. Simple adiabatic models fail to fit the emissivity
variations.
\end{abstract}

\begin{keywords}
galaxies: jets -- radio continuum:galaxies -- magnetic fields -- polarization --
MHD
\end{keywords}

%------------------------ ****** INTRODUCTION ****** ------------------------%
\section{Introduction}
\label{intro}

It was first recognised by \cite{FR74} that the ways in which extragalactic jets
dissipate energy to produce observable radiation differ for high- and
low-luminosity sources. Jets in low-luminosity (FR\,I) sources are bright close
to the nucleus of the parent galaxy, whereas those in powerful (FR\,II) sources
are relatively faint until their terminal hot-spots. It rapidly became accepted
that FR\,I jets must decelerate by entrainment of the surrounding IGM
\citep{Baa80,Beg82,Bic84,Bic86,DeY96,DeY04,RHCJ99,RH00} or by injection of mass
lost by stars within the jet volume \citep{Phi83,Kom94,BLK96}.

More recently, evidence has accumulated that FR\,I jets are initially
relativistic and decelerate on kpc scales. FR\,I sources are thought to be the
side-on counterparts of BL Lac objects, in which relativistic motion on parsec
scales is well-established \citep{UP95}.  Proper motions corresponding to speeds
comparable with and in some cases exceeding $c$ have been measured on
milliarcsecond scales in several FR\,I jets \citep{Giov01} and on arcsecond
scales in M\,87 and Cen\,A \citep{Biretta95,Hard03}. In FR\,I sources (as in
FR\,IIs), the lobe containing the main (brighter) jet is less depolarized than
the counter-jet lobe \citep{Morganti97}. This can be explained as an effect of
Faraday rotation in the surrounding halo of hot plasma if the main jet points
toward the observer, suggesting that the brightness asymmetry is caused by
Doppler beaming \citep{Laing88}. The decrease of this asymmetry with distance
from the nucleus \citep{LPdRF} indicates deceleration.

The present paper is the third in a series devoted to modelling the jets in
FR\,I radio galaxies.  We assume that the jets are intrinsically symmetrical,
axisymmetric, relativistic flows and we parameterize their geometry and the
three-dimensional variations of their velocity, emissivity and magnetic-field
structure. We then compute the brightness distributions in total intensity and
linear polarization. By fitting simultaneously to deep, high-resolution radio
images in Stokes $I$, $Q$ and $U$, we can optimize the model parameters. The
fits are empirical, and make as few assumptions as possible about the
(poorly-known) internal physics of the jets. The technique was originally
developed to model the radio galaxy 3C\,31 by \citet[hereafter LB]{LB02a} and
was then slightly revised and applied to B2\,0326+39 and B2\,1553+24 by
\citet[hereafter CL]{CL}.  We now present a model for the jets in the well-known
nearby radio galaxy NGC\,315.  This source has a huge angular and physical size
\citep{Brid76} and, as we shall show, the angular scales on which flaring,
recollimation and deceleration of the jets take place are much larger than in
the sources we have studied so far. Our results for NGC\,315 therefore give a more
detailed picture of the initial propagation of an FR\,I jet.

In order to improve on our empirical models, we need to understand the energy
gain and loss processes affecting the ultrarelativistic particles which produce
the observed synchrotron radiation. A self-consistent, axisymmetric adiabatic
model fails to fit either the total intensity or the polarization distributions
in the jets of 3C\,31 within 5\,kpc of the nucleus \citep{LB04}. This suggests
that injection of relativistic particles and/or amplification of the magnetic
field are required, which is not surprising in view of the detection of
cospatial X-ray synchrotron emission \citep{Hard02}. Further out in 3C\,31, the
adiabatic model gives a tolerable fit. A simple, quasi-one-dimensional analysis
(LB, CL) is adequate to assess whether fitting of more elaborate models is
worthwhile, and we apply this to NGC\,315.

Given a kinematic model for the jets and estimates of the external density and
pressure from X-ray observations, we can apply conservation of mass, momentum
and energy to deduce the variations of internal pressure, density, Mach number
and entrainment rate with distance from the nucleus \citep{LB02b}.

In Section~\ref{obs}, we introduce NGC\,315 and briefly summarize the VLA
observations. Our modelling technique is outlined in Section~\ref{model},
emphasizing the (small) differences from the earlier work of CL. The observed
and model brightness distributions are compared in Section~\ref{results} and the
derived geometry, velocity, emissivity and field distributions are presented in
Section~\ref{physical}.  We summarize our conclusions and outline our future
programme in Section~\ref{ssfw}.

We adopt a concordance cosmology with Hubble constant, $H_0$ =
70\,$\rm{km\,s^{-1}\,Mpc^{-1}}$, $\Omega_\Lambda = 0.7$ and $\Omega_M = 0.3$,
although only the choice of $H_0$ is significant at the distance of NGC\,315.

%--------------------------- ****** DATA ****** -----------------------------%
\section{Observations and images}
\label{obs}

\subsection{NGC\,315}
\label{ngc315}

\begin{figure}
\includegraphics[width=8.5cm]{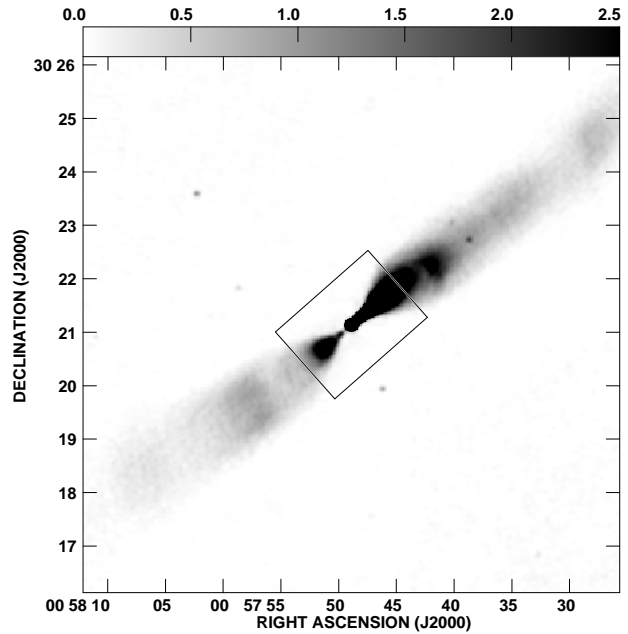}
\caption{A grey-scale of the large-scale structure of the jets in NGC\,315 at a
  resolution of 5.5\,arcsec FWHM. The image is the mean of four at different
  frequencies between 1.365 and 1.665\,GHz and is described in
  Laing et al.\ (in preparation). The grey-scale range is 0 -- 2.5\,mJy\,(beam area)$^{-1}$
  as shown by the labelled wedge and the modelled area is indicated by the box.  
\label{fig:modelarea}}
\end{figure}

Our modelling technique requires that both radio jets are: 
\begin{enumerate}
\item detectable with good signal-to-noise ratio in total intensity and linear
  polarization;
\item straight and antiparallel;
\item separable from any surrounding lobe emission and
\item asymmetric, in the sense that their jet/counter-jet intensity ratio is
significantly larger than unity over a significant area.
\end{enumerate}
The nearby giant elliptical galaxy NGC\,315, whose large-scale radio structure
was first imaged by \citet{Brid76}, is one of the brightest sources satisfying
these criteria.  The galaxy has a redshift of 0.01648 \citep{Trager}, giving a
scale of 0.335\,kpc arcsec$^{-1}$ for our adopted cosmology.  The overall extent
of the radio source is approximately 3500\,arcsec, or about 1200\,kpc in
projection, but the area to be modelled (see Fig.~\ref{fig:modelarea}) is
limited in extent by the slight bends in the jet at roughly 70 arcsec (23\,kpc
in projection) from the nucleus (see Section~\ref{outline-method}).   

 High-resolution images of the jets on kpc scales were presented by
\citet{Brid79}, \citet{Fom80} and \citet{Venturi93} and on pc scales by
\citet{Linfield81}, \citet{Venturi93}, \citet{Cotton99} and \citet{Xu00}.  More
comprehensive references to radio, optical and X-ray observations of NGC\,315
are given by Laing et al.\ (in preparation).

\subsection{Observations and data reduction}

The observations and their reduction are also described in detail by 
Laing et al.\ (in preparation).  We fit to images made using the 5\,GHz dataset from that
paper at 2.35 and 0.40\,arcsec FWHM [the higher-resolution images are also
discussed and compared with {\sl Chandra} observations by Worrall et al.\ (in
preparation)].  The dataset includes long observations in all four
configurations of the VLA and provides excellent sampling of spatial scales
between 0.4 and 300\,arcsec. The resolutions and noise levels are given in
Table~\ref{noise}.

\begin{center}
\begin{table}
\caption{Image resolutions and noise levels. $\sigma_I$ is the
  off-source noise level on the $I$ image; $\sigma_P$ the average of
  the noise levels for $Q$ and $U$.
\label{noise}}
\begin{tabular}{lcc}
\hline
 FWHM  &\multicolumn{2}{c|}{rms noise level} \\
  (arcsec)  &\multicolumn{2}{c|}{[$\mu$Jy\,(beam area)$^{-1}$]} \\
            &$\sigma_I$&$\sigma_P$ \\
\hline
   2.35   & 10.0& 7.5 \\      
   0.40   & 12.5& 6.8 \\
\hline
\end{tabular}
\end{table}
\end{center}

The accuracy of linear polarization measurements is critical to our technique.
The {\bf E}-vector position angles have been corrected to zero wavelength using
an image of rotation measure derived from 5-frequency observations at a
resolution of 5.5\,arcsec FWHM between 1.365 and 5\,GHz
(Laing et al., in preparation). Significant corrections are required for the mean Faraday
rotation ($-76$\,rad\,m$^{-2}$) and a linear gradient along the jet, both
probably of Galactic origin. The fitting errors are typically
$\la$2.5\,rad\,m$^{-2}$ in the area of interest, corresponding to position angle
rotation $\la$0.5$^\circ$ at 5\,GHz.  The observed fluctuations of rotation
measure on scales of a few beamwidths are very small ($\approx$1 --
2\,rad\,m$^{-2}$) and residual depolarization is undetectable, so errors in {\bf
E}-vector position angle and degree of polarization $p = (Q^2+U^2)^{1/2}/I$ due
to changes in Faraday rotation across the 5.5\,arcsec beam should be
negligible. Observations at two centre frequencies (4.860 and 4.985\,GHz) were
combined to make the final dataset (Laing et al., in preparation) but the error in {\bf
E}-vector position angle due to this effect is $<$1$^\circ$. Depolarization due
to rotation across the maximum bandwidth of 50\,MHz is also negligible. The
modelling described later is not affected by Ricean bias \citep{WK}, as we fit
to Stokes $I$, $Q$ and $U$ directly, but our plots of the degree of polarization
include a first-order correction for this effect. We plot the direction of the
{\sl apparent magnetic field}, rotated from the zero-wavelength {\bf E}-vector
position angles by 90$^\circ$.

%-------------------------- ****** MODEL ****** ----------------------------%
\section{The model}
\label{model}

\subsection{Assumptions}

Our fundamental assumptions are those of LB and CL:
\begin{enumerate}
\item The jets may be modelled as intrinsically symmetrical, antiparallel,
axisymmetric, stationary, laminar flows. The real flow is clearly much more
complicated, but all that is necessary for our technique to work is that an average
over a sufficiently large volume results in statistically identical rest-frame
emission from the main and counter-jets. In poorly resolved cases, some care is
required to distinguish between intrinsic variations of physical parameters and
local fluctuations such as knots and filaments, as we will discuss.
\item The jets contain relativistic particles with an energy spectrum $n(E)dE = n_0
E^{-(2\alpha+1)}dE$ (corresponding to a frequency spectral index $\alpha$) with
an isotropic pitch-angle distribution. The maximum degree of linear polarization
is then $p_0 = (3\alpha+3)/(3\alpha+5)$. We use $\alpha = 0.55$, the mean
spectral index for the modelled region between 1.365 and 5\,GHz (Laing et al., in preparation).
\item The magnetic field is tangled on small scales, but anisotropic (the
reasons for taking the field to be of this form are discussed by
\citealt{Laing81}, \citealt{BBR} and LB; see also
Section~\ref{field-structure}).
\item The emission is optically thin everywhere. The core, which is unresolved
  in our VLA observations and partially optically thick, is not modelled. It is
  represented as a point source with the correct flux density.  
\end{enumerate}
We define $\beta = v/c$, where $v$ is the flow velocity. $\Gamma =
(1-\beta^2)^{-1/2}$ is the bulk Lorentz factor and $\theta$ is the angle between
the jet axis and the line of sight.

\subsection{Outline of the method}
\label{outline-method}

The jet/counter-jet intensity ratio for intrinsically symmetrical, relativistic
jets depends on the product $\beta\cos\theta$, so the angle to the line of sight
and velocity cannot be decoupled by observations of $I$ alone.  The basis of our
technique is that relativistic aberration causes the approaching and receding
jets to appear different not only in total intensity, {\sl but also in linear
polarization}. We use these polarization differences as independent constraints
in order to break the degeneracy between $\beta$ and $\theta$.  An outline of
our procedure is as follows.
\begin{enumerate}
\item Develop a parameterized description of the geometry, velocity, emissivity
  and magnetic-field ordering.
\item Calculate the synchrotron emission in $I$, $Q$ and $U$ from the model jets
  by numerical integration, taking account of relativistic aberration.
\item Convolve and compare with deep VLA images, using $\chi^2$ to measure the
  quality of the fit.
\item Optimize the model parameters using the downhill simplex method.
\end{enumerate} 
The details of the calculations are fully described by LB and CL.

The assumption of axisymmetry is clearly violated in NGC\,315 at
$\approx$70\,arcsec from the nucleus. Both jets bend clockwise by
$\approx$5$^\circ$ (in projection) at this distance, but we do not know the
magnitude of any associated bends in the orthogonal direction. Our modelling
could be extended to larger distances provided that the jets are indeed
antiparallel and intrinsically identical after the bends. This seems plausible
from their appearance in projection and we could fit independently for the angle
to the line of sight after the bend if necessary.  The brightness and
polarization structures of the jets are qualitatively consistent with an
extrapolation of the fitted model described below, so it is likely that this
approach would succeed. It adds significant complexity to the modelling
procedure, however, and also requires careful justification of the assumption of
intrinsic symmetry at large distances. For these reasons, we defer this analysis
to a later paper.

\subsection{Functional forms for geometry, velocity, magnetic field and
  emissivity}

The functional forms used to parameterize the geometry, velocity, emissivity and
magnetic field are very similar to those described by CL, which are in
turn an evolution from those introduced by LB. The earlier papers
discuss the motivation of these expressions in detail; here we just summarize their
forms for completeness and to highlight a few differences from our earlier work.
 
\subsubsection{Geometry}
\label{geom-parms}

\begin{figure}
\includegraphics[width=8.5cm]{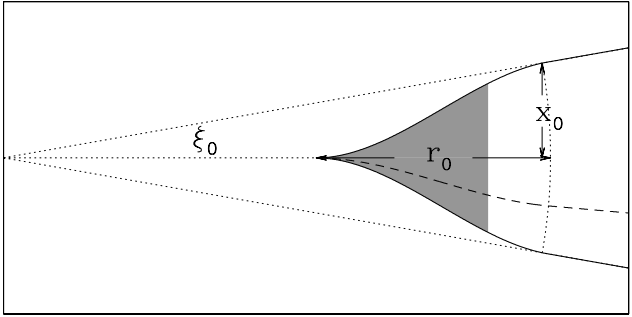}
\caption{A sketch of the assumed jet geometry, defining the quantities $\xi_0$,
  $r_0$ and $x_0$ used in Table~\ref{tab:results}. For NGC\,315, the modelled
  area (indicated schematically by shading) is entirely within the flaring
  region. The full line indicates the edge of the jet ($s = 1$) and the dashed
  line the $s = 0.5$ streamline.
\label{fig:geom}}
\end{figure}
The assumed geometry, sketched in Fig.~\ref{fig:geom} is derived from a fit to
the outer isophote of the jet emission. It is identical to that used by CL, but
in this case we model only the {\em flaring region} where the jet expands
rapidly and then recollimates. NGC\,315 also shows a conical {\em outer region}
of constant (very small) opening angle, as seen in other FR\,I jets, but this is
beyond the bends which limit our ability to fit an axisymmetric model directly
to the observed data.

We define $z$ to be the distance from the nucleus along the axis and $\xi_0$ to
be the half-opening angle of the jet in the outer region.  As in CL, our assumed
flow streamlines are a family of curves one of whose members is defined by the
outer boundary of the jet.  We use a streamline coordinate system $(\rho,s)$
where the streamline index $s$ is constant for a given streamline and $\rho$
increases monotonically with distance along it. In the outer region, the
streamlines are assumed to be radial from a point on the axis with $z = -A$,
where $A = x_0/\sin\xi_0 - r_0$. For a streamline which makes an angle $\xi$
with the jet axis in the outer region, we define $s = \xi/\xi_0$, so $s = 0$
on-axis and $s = 1$ at the edge (Fig.~\ref{fig:geom}).  The distance of a
streamline from the jet axis in the flaring region is:
\begin{eqnarray*}
x(z,s) & = & a_2(s) z^2 + a_3(s) z^3 \\
\end{eqnarray*}
The curve is constrained to fit the jet boundary for $s = 1$ and the
coefficients are determined by the conditions that $x(s)$ and its first
derivative are continuous at the boundary between the regions. $\rho$ is defined
by:
\begin{eqnarray*}
\rho & = & \frac{zr_0}{(r_0 + A) \cos (\xi_0s) - A} \hspace{1.0cm}
\rho < r_{0} \\
\end{eqnarray*}
and varies monotonically along a streamline from 0 at the nucleus to $r_0$ at
the spherical surface defining the end of the flaring region.

\subsubsection{Velocity}
\label{vel-parms}

The form of the velocity field $\beta(\rho,s)$ is the same as that used by
CL. The on-axis velocity profile is divided into three parts:
\begin{enumerate}
\item roughly constant, with a high velocity $\beta_1$, close to the nucleus;
\item a linear decrease and
\item roughly constant but with a low velocity $\beta_0$ at large distances.
\end{enumerate}
[the velocities in (i) and (iii) are not quite constant because transitions
between regions are smoothed to avoid numerical problems].  The profile is
defined by four free parameters: the distances of the two boundaries separating
the three regions, $\rho_{\rm v_1}$ and $\rho_{\rm v_0}$, and the characteristic
inner and outer velocities $\beta_1$ and $\beta_0$. Off-axis, the velocity is
calculated using the same expressions but with inner and outer velocities
$\beta_1\exp(-s^2\ln v_1)$ and $\beta_0\exp(-s^2\ln v_0)$, i.e.\
with a truncated Gaussian transverse profile falling to fractional velocities
$v_1$ and $v_0$ at the edge of the jet in the inner and outer regions,
respectively.

The full functional forms for the velocity field $\beta(\rho,s)$ are given in
Table~\ref{tab:param}.

\subsubsection{Magnetic field}
\label{field-parms}

We define the rms components of the magnetic field to be $\langle
B_l^2\rangle^{1/2}$ (longitudinal, parallel to a streamline), $\langle
B_r^2\rangle^{1/2}$(radial, orthogonal to the streamline and outwards from the
jet axis) and $\langle B_t^2\rangle^{1/2}$ (toroidal, orthogonal to the
streamline in an azimuthal direction).  The magnetic-field structure is
parameterized by the ratio of rms radial/toroidal field, $j(\rho,s) = \langle
B_r^2\rangle^{1/2}/\langle B_t^2\rangle^{1/2}$ and the longitudinal/toroidal
ratio $k(\rho,s) =\langle B_l^2\rangle^{1/2}/\langle B_t^2\rangle^{1/2}$.  The
parameterization is similar to that used by CL, but we also allow variation of
the field component ratios with $s$ (cf.\ LB) in order to improve the fit to the
variation of polarization across the jets, which is better resolved in NGC\,315
than in the sources modelled by CL.  For each of the axial ($s = 0$) and edge
($s = 1$) streamlines, the field ratios have constant values for $\rho <
\rho_{\rm B_1}$ and $\rho > \rho_{\rm B_0}$, with linear interpolation between
them for $\rho_{\rm B_1} \leq \rho \leq \rho_{\rm B_0}$. For intermediate 
streamlines, we then interpolate linearly between the ratios for $s = 0$ and $s
= 1$.

The functional forms assumed for the field ratios are again given in
Table~\ref{tab:param}.

\begin{figure*}
\includegraphics[width=17cm]{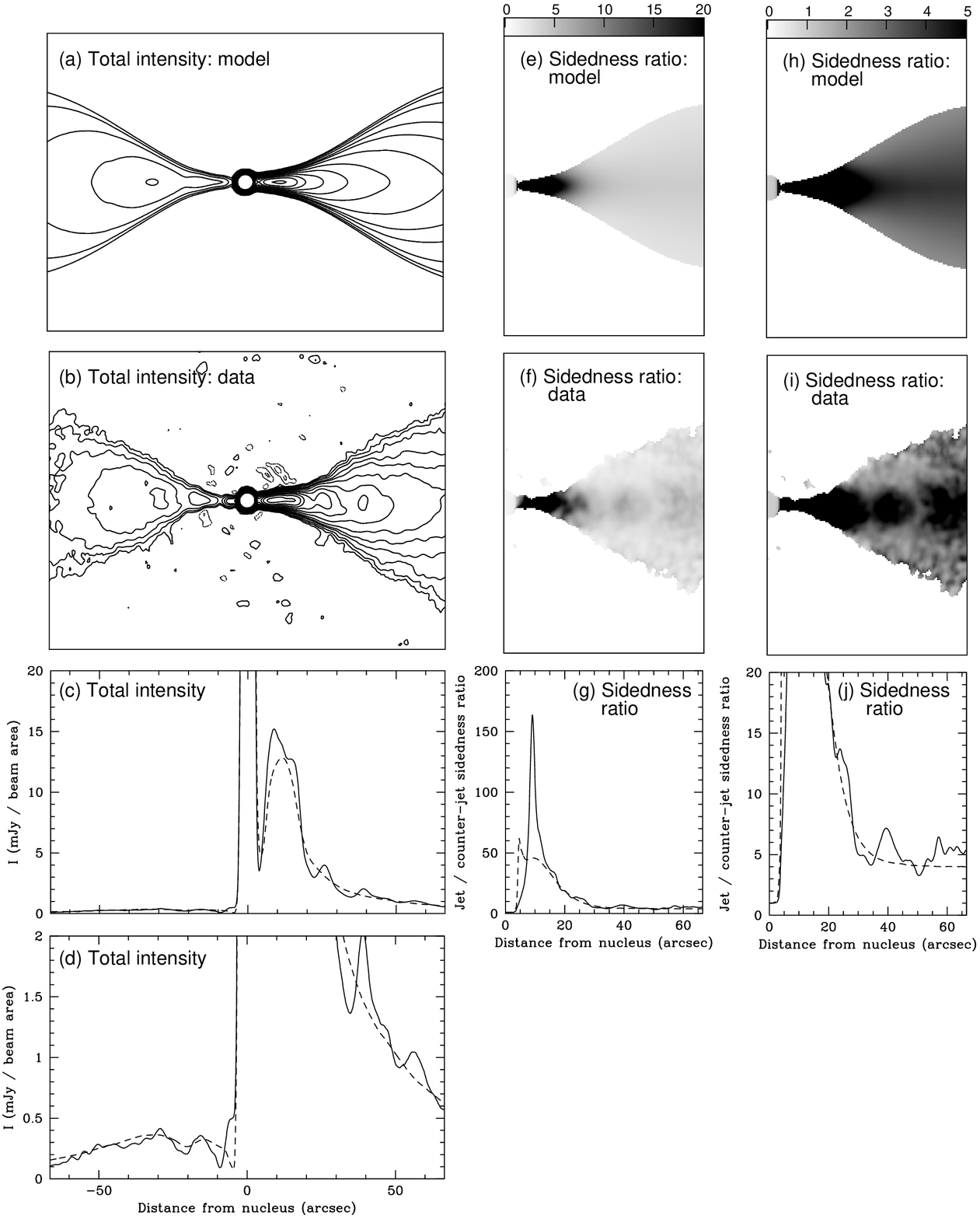}
\caption{A comparison of the model and data in total intensity at 2.35 arcsec
resolution for the inner 66.5 arcsec of each jet. (a) model contours; (b)
observed contours. The contour levels in panels (a) and (b) are $-2$, $-1$, 1,
2, 4, 8, 12, 20, 36, 60, 100, 160, 250, 400, 600, 1000 $\times$
30\,$\mu$Jy\,(beam area)$^{-1}$. (c) and (d) $I$ profiles along the jet axis for
the data (solid line) and model (dashed line). (e) and (f) Grey-scales of
sidedness ratio, obtained by dividing total-intensity images by copies of
themselves rotated by 180$^\circ$ about the core. The ratio is in the sense main
jet / counter-jet. (e) model; (f) data. (g) profile of sidedness ratio
along the jet axis for the data (solid line) and model (dashed line). (h) -- (j)
as (e) -- (g), but with smaller grey-scale and profile ranges to emphasise the
variation of sidedness in the outer parts of the modelled region.
\label{fig:ilo}}
\end{figure*}

\subsubsection{Emissivity}
\label{emiss-parms}

We write the proper emissivity as $\epsilon(\rho,s) h(\rho, s)$, where
$\epsilon$ is the emissivity in $I$ for a magnetic field $B = \langle B_l^2 +
B_r^2 + B_t^2\rangle^{1/2}$ perpendicular to the line of sight. $h$ depends on
field geometry: for $I$, $0 \leq h \leq 1$ and for $Q$ and $U$ $-p_0 \leq h \leq
+p_0$.  We refer to $\epsilon$, loosely, as `the emissivity'. For a given
spectral index, it is a function only of the rms total magnetic field and the
normalizing constant of the particle energy distribution, $\epsilon \propto n_0
B^{1+\alpha}$.

The on-axis emissivity profile consists of five regions, each with a power-law
variation of $\epsilon$ with $\rho$. The profile for NGC\,315 is consistent with
a continuous variation of emissivity, so we enforce continuity everywhere ($g =
1$ in the notation of CL).  We use the expressions given by CL, but change the
notation slightly in the interests of greater clarity.  The profile is defined
by five power-law indices and four boundary positions.  As explained in more
detail in Section~\ref{results:emiss}, there are three primary emissivity
regions with indices $E_{\rm in}$, $E_{\rm mid}$ and $E_{\rm out}$. The
remaining two regions are introduced to model the transitions between them. The
first (slope $E_{\rm rise}$) replaces the discontinuous increase in emissivity
used to model the initial brightening of the jets in 3C\,31 and B2\,0326+39 (LB,
CL) because the equivalent structure in NGC\,315 is better resolved. The second
(slope $E_{\rm fall}$), which models an abrupt decrease in emissivity, is
equivalent to $E_3$ in the models of B2\,0326+39 and B2\,1553+24 (CL).

Off-axis, the profile is multiplied by a factor $\exp[-s^2\ln\bar{e}(\rho)]$, so
that $\bar{e}(\rho)$ is the fractional value of the emissivity at the jet edge.
$\bar{e}(\rho)$ has a constant value $e_0$ for $\rho > \rho_{\rm e_3}$ and
varies linearly through regions 2 and 3 from $e_1$ at $\rho_{\rm e_1}$ to $e_0$
at $\rho_{\rm e_3}$. For $\rho < \rho_{\rm e_1}$, the jet is too narrow for our
data to constrain any transverse profile and we set $\bar{e}(\rho) = 1$.

The full description of the emissivity distribution $\epsilon(\rho,s)$ is
given in Table~\ref{tab:param}. 

\begin{table*}
\caption{Functional forms of the velocity $\beta$, emissivity
$\epsilon$, radial/toroidal and longitudinal/toroidal magnetic-field
ratios $j$ and $k$ in the streamline coordinate system
$(\rho,s)$. Column 4 lists the parameters which may be optimized, for
comparison with Table~\ref{tab:results}.}
\begin{minipage}{160mm}
\begin{center}
\begin{tabular}{llcl}
\hline
&&&\\
Quantity & Functional form & Range & Free parameters \\
&&&\\
\hline
&&&\\
\multicolumn{4}{c}{Velocity field\footnote{Note that the constants $c_1$,
  $c_2$, $c_3$ and $c_4$ 
are defined by the values of the free parameters and the conditions that the 
velocity and acceleration are continuous at the two boundaries.}}\\ 
&&&\\
$\beta(\rho,s)$ &  $\beta_{1} - \left[\frac{\beta_{1}\exp(-s^2\ln v_1) - \beta_{0}\exp(-s^2\ln v_0)}{10}\right]\exp[c_{1}(\rho - \rho_{\rm v_1})]$ 
& $\rho < \rho_{\rm v_1}$ & Distances $\rho_{\rm v_1}$, $\rho_{\rm
  v_0}$ \\
&&&\\
& $c_{2} + c_{3}\rho$ 
& $\rho_{\rm v_1} \le \rho \le \rho_{\rm v_0}$ & Velocities $\beta_1$,
$\beta_0$\\
&&&\\
& $\beta_{0} + \left[\frac{\beta_{1}\exp(-s^2\ln v_1) - \beta_{0}\exp(-s^2\ln v_0)}{10}\right]\exp[c_{4}(\rho_{\rm v_0} - \rho)]$ 
& $\rho > \rho_{\rm v_0}$\footnote{There is a typographical error in the
  equivalent expression in CL, which should be the same as that given here} & Fractional edge velocities $v_1$, $v_0$\\
&&&\\
&&&\\
\multicolumn{4}{c}{Emissivity}\\
&&&\\
$\epsilon(\rho,s)$ & $~~~\left(\frac{\rho}{\rho_{\rm e_1}}\right)^{-E_{\rm in}}$ 
& $\rho \le \rho_{\rm e_1}$ & Distances $\rho_{\rm e_1}$, $\rho_{\rm
  e_2}$, $\rho_{\rm e_3}$, $\rho_{\rm e_4}$ \\
&&&\\
&~~~$\left(\frac{\rho}{\rho_{\rm e_1}}\right)^{-E_{\rm rise}}
\exp\left[-s^2\ln\left(e_1 + (e_0 - e_1)\left(\frac{\rho - \rho_{\rm e_1}}{\rho_{\rm e_2} - \rho_{\rm e_1}}\right)\right)\right]$ 
& $\rho_{\rm e_1} < \rho \le \rho_{\rm e_2}$ & Indices $E_{\rm in}$, $E_{\rm rise}$, $E_{\rm mid}$, $E_{\rm fall}$, $E_{\rm out}$  \\
&&&\\
& $d_1\left(\frac{\rho}{\rho_{\rm e_2}}\right)^{-E_{\rm mid}}
\exp\left[-s^2\ln\left(e_1 + (e_0 - e_1)\left(\frac{\rho - \rho_{\rm e_1}}{\rho_{\rm e_2} - \rho_{\rm e_1}}\right)\right)\right]$
& $\rho_{\rm e_2} < \rho \le \rho_{\rm e_3}$ &Fractional edge emissivities $e_1$, $e_0$\\
&&&\\
&$d_2 \left(\frac{\rho}{\rho_{\rm e_3}}\right)^{-E_{\rm fall}} \exp(-s^2\ln e_0)$ 
& $\rho_{\rm e_3} < \rho \le \rho_{\rm e_4}$ & \\
&&&\\
&$d_3 \left(\frac{\rho}{\rho_{\rm e_4}}\right)^{-E_{\rm out}} \exp(-s^2\ln e_0)$ 
& $\rho > \rho_{\rm e_4}$ &\\
&&&\\
&&&\\
\multicolumn{4}{c}{Radial/toroidal field ratio}\\
&&&\\
$j(\rho,s)$ &  $j_{1}^{\rm axis} + s(j_{1}^{\rm edge}- j_{1}^{\rm axis})$ 
& $\rho \le \rho_{\rm B_1}$ & Distances $\rho_{\rm B_1}$, $\rho_{\rm B_0}$\\
&&&\\
&$j^{\rm axis} + s(j^{\rm edge}- j^{\rm axis})$ 
& $\rho_{\rm B_1} < \rho < \rho_{\rm B_0}$   & Ratios $j_1^{\rm edge}$, $j_0^{\rm
  edge}$, $j_1^{\rm axis}$, $j_0^{\rm axis}$\\
&where~~$j^{\rm axis} = j_{1}^{\rm axis} + (j_{0}^{\rm axis} - j_{1}^{\rm axis})\left(\frac{\rho - \rho_{\rm B_1}} {\rho_{\rm B_0} - \rho_{\rm B_1}}\right)$ 
& &\\
&~~~~~~~~~$j^{\rm edge} = j_{1}^{\rm edge} + (j_{0}^{\rm edge} - j_{1}^{\rm axis})\left(\frac{\rho - \rho_{\rm B_1}} {\rho_{\rm B_0} - \rho_{\rm B_1}}\right)$ 
& &\\
&&&\\
&$j_{0}^{\rm axis} + s(j_{0}^{\rm edge}- j_{0}^{\rm axis})$ 
& $\rho \ge \rho_{\rm B_0}$ &\\
&&&\\
&&&\\
\multicolumn{4}{c}{Longitudinal/toroidal field ratio}\\
&&&\\
$k(\rho,s)$ &  $k_{1}^{\rm axis} + s(k_{1}^{\rm edge}- k_{1}^{\rm axis})$  
& $\rho \le \rho_{\rm B_1}$ & Ratios  $k_1^{\rm edge}$, $k_0^{\rm
  edge}$, $k_1^{\rm axis}$, $k_0^{\rm axis}$ \\
&&&\\
&$k^{\rm axis} + s(k^{\rm edge}- k^{\rm axis})$ 
& $\rho_{\rm B_1} < \rho < \rho_{\rm B_0}$   &\\
&where~~$k^{\rm axis} = k_{1}^{\rm axis} + (k_{0}^{\rm axis} - k_{1}^{\rm axis})\left(\frac{\rho - \rho_{\rm B_1}} {\rho_{\rm B_0} - \rho_{\rm B_1}}\right)$ 
& &\\
&~~~~~~~~~$k^{\rm edge} = k_{1}^{\rm edge} + (k_{0}^{\rm edge} - k_{1}^{\rm edge})\left(\frac{\rho - \rho_{\rm B_1}} {\rho_{\rm B_0} - \rho_{\rm B_1}}\right)$ 
& &\\
&&&\\
& $k_{0}^{\rm axis} + s(k_{0}^{\rm edge}- k_{0}^{\rm axis})$ 
& $\rho \ge \rho_{\rm B_0}$ & \\
&&&\\
\hline
\end{tabular}
\end{center}
\end{minipage}
\label{tab:param}
\end{table*}

\subsection{Fitting, model parameters and errors}
\label{fit-details}

The noise level for the calculation of $\chi^2$ for $I$ is estimated as
$1/\sqrt{2}$ times the rms of the difference between the image and a copy of
itself reflected across the jet axis. For linear polarization, the same level is
used for both $Q$ and $U$. This is the mean of $1/\sqrt{2}$ times the rms of the
difference image for $Q$ and the summed image for $U$, since the latter is
antisymmetric under reflection for an axisymmetric model. This prescription is
identical to that used by LB and CL. The region immediately around the core is
excluded from the fit.  $\chi^2$ is calculated using the high-resolution
(0.4\,arcsec) images where they have adequate signal-to-noise ratio (0.9 --
17\,arcsec from the nucleus) and the lower-resolution (2.35\,arcsec) images
elsewhere (17 -- 67\,arcsec). The same fitting regions are used for total
intensity and linear polarization and $\chi^2$ values for $I$, $Q$ and $U$ are
summed over grids chosen so that all points are independent.

\begin{figure}
\includegraphics[width=8.5cm]{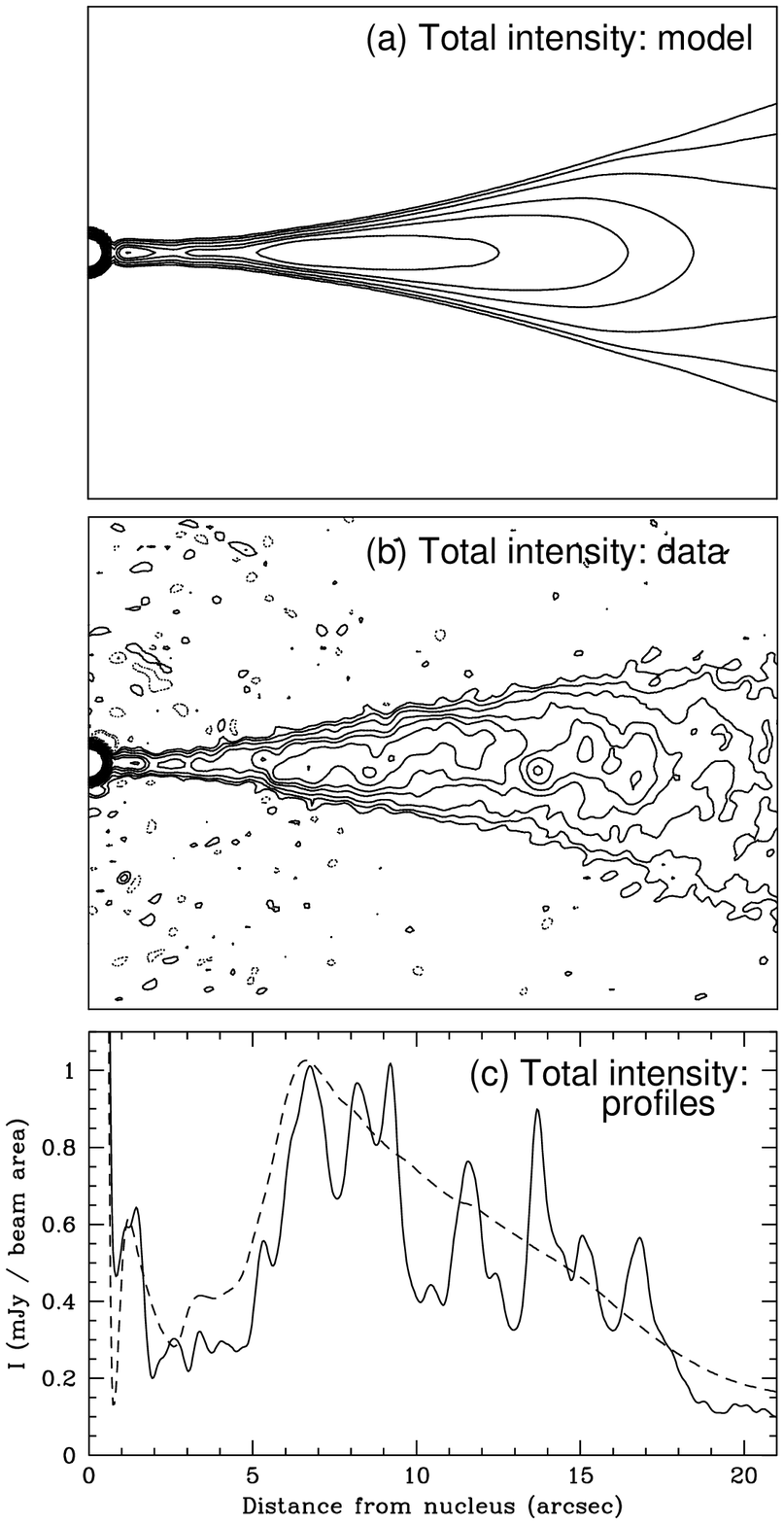}
\caption{A comparison of the model and data in total intensity at 0.4 arcsec
resolution for the inner 21\,arcsec of the main jet. (a) model contours; (b)
observed contours. The contour levels in panels (a) and (b) are $-2$, $-1$, 1,
2, 4, 8, 12, 20, 36, 60, 100, 160, 250, 400, 600, 1000 $\times$ 30\,$\mu$Jy/beam
area. (c) $I$ profile along the jet axis for the data (solid line) and model
(dashed line).  
\label{fig:ihi}}
\end{figure}

\begin{figure}
\includegraphics[width=8.5cm]{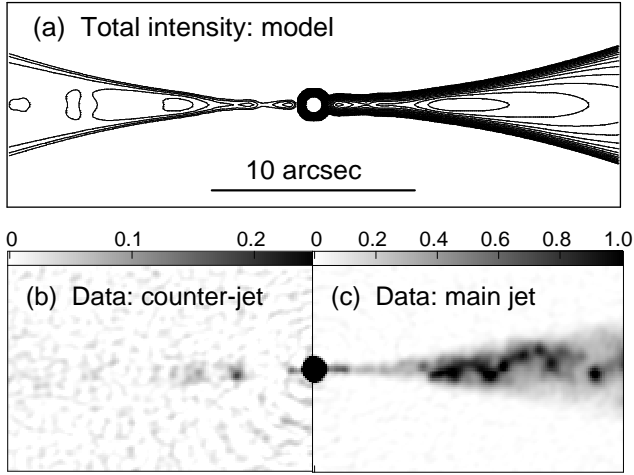}
\caption{A comparison of the model and data in total intensity at 0.4 arcsec
resolution for the inner $\pm$30\,arcsec. (a) Model contours. The levels are
logarithmic, increasing by factors of $\sqrt{2}$ from 4.24 -- 543\,$\mu$Jy (beam
area)$^{-1}$. (b) Grey-scale of observed emission from the counter-jet. The
range is 0 -- 0.25\,mJy (beam area)$^{-1}$. (c) as in panel (b), but for the
main jet with a range of 0 -- 1\,mJy (beam area)$^{-1}$. This display emphasizes
the fine-scale structure in both jets which cannot be described by the model,
but shows that the gross features of the predicted and observed brightness
distributions are very similar.
\label{fig:ismall}}
\end{figure}

\begin{figure}
\includegraphics[width=8.5cm]{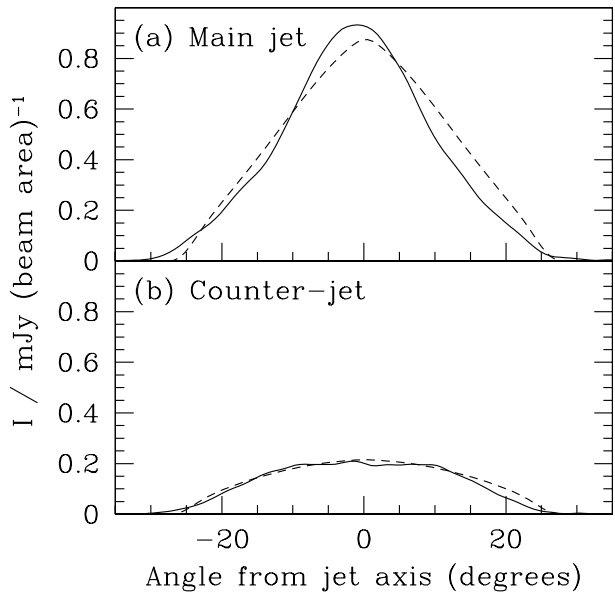}
\caption{Average transverse profiles of total intensity at a resolution of
  2.35\,arcsec. The profiles are generated by averaging along radii between 45
  and 66.5\,arcsec from the nucleus and plotting against angle from the jet
  axis. The data are represented by the full line and the model by the dashed
  line. (a) main jet; (b) counter-jet. The jets are not perfectly straight, so
  there is a slight offset between the observed and model profiles over this
  range of distances.
\label{fig:transi}}
\end{figure}

\begin{figure}
\includegraphics[width=7.5cm]{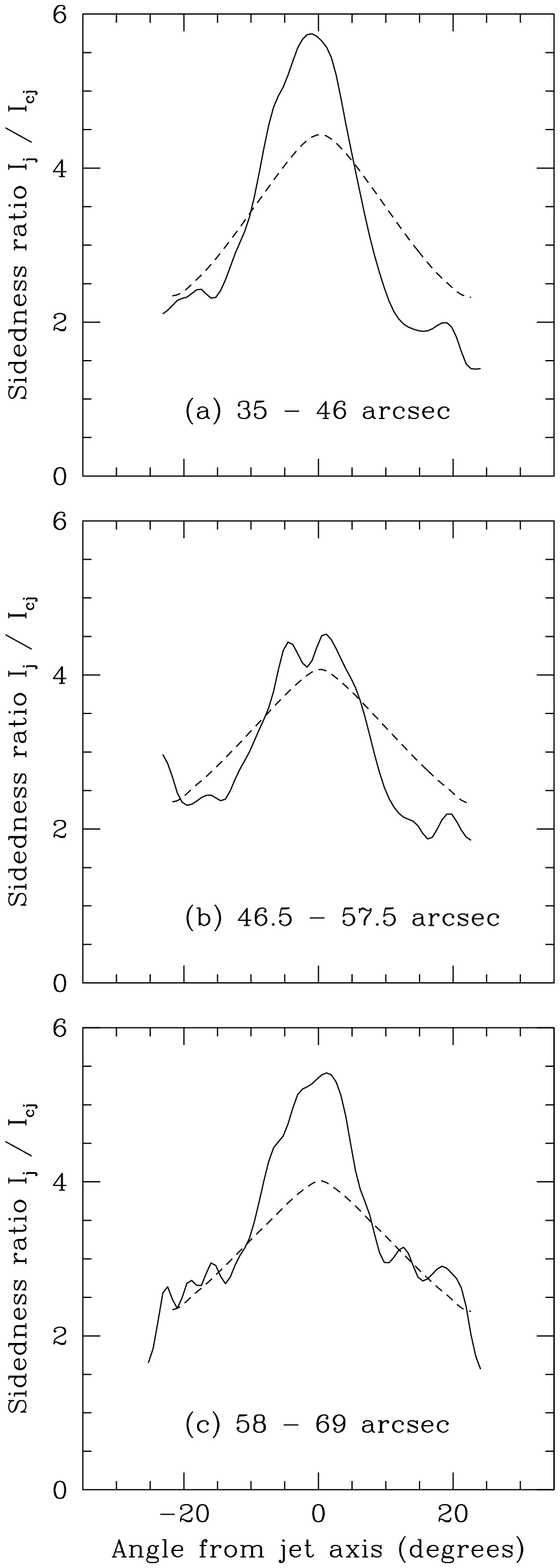}
\caption{Average transverse profiles of the jet/counter-jet sidedness ratio
  $I_{\rm j}/I_{\rm cj}$ at a resolution of 2.35\,arcsec. The profiles are
  generated by averaging along radii from the nucleus and plotting against angle
  from the jet axis. The data are represented by the full line and the model by
  the dashed line. (a) 35 -- 46\,arcsec; (b) 46.5 -- 57.5\,arcsec; (c) 58 --
  69\,arcsec. 
\label{fig:transside}}
\end{figure}

We derive rough uncertainties, as in LB and CL, by varying individual parameters
until the increase in $\chi^2$ corresponds to the formal 99\% confidence level
for independent Gaussian errors. These estimates are crude (they neglect
coupling between parameters), but in practice give a good representation of the
range of qualitatively reasonable models.  The number of independent points
(1414 in each of 3 Stokes parameters) is sufficiently large that we are
confident in the main features of the model.

\section{Comparison between models and data}
\label{results}

The quality of the fits is extremely good and the reduced $\chi^2 = 1.27$ for
4242 independent points.

\subsection{Total intensity}

The observed and modelled total intensities and jet/counter-jet sidedness ratios
are displayed in Figs~\ref{fig:ilo} -- \ref{fig:transside}.  Fig.~\ref{fig:ilo}
shows the inner 66.5\,arcsec of both jets at a resolution of 2.35\,arcsec FWHM
and Fig.~\ref{fig:ihi} shows the inner 21\,arcsec of the main jet at 0.4\,arcsec
FWHM.  Fig.~\ref{fig:ismall} shows grey-scales of the observed total intensity
for the inner 30\,arcsec of both jets at the same resolution, but with levels
chosen to emphasize the fine-scale structure.  These are compared with contours
of the model $I$. Figs~\ref{fig:transi} and \ref{fig:transside} show the average
transverse profiles of total intensity and jet/counter-jet sidedness ratio over
areas where the latter is nearly independent of distance. [The small offset
between the observed and model profiles visible in both of these figures is
caused by slight bends in the jets between small and large scales.]

The following total-intensity features are described accurately by the model:
\begin{enumerate}
\item The jets are initially well collimated and flare to a projected opening
  angle of $\approx 30^\circ$ at a distance of 20\,arcsec from the nucleus
  (Figs~\ref{fig:ilo}a and b and \ref{fig:ihi}a and b).
\item Both jets are faint and narrow within 5\,arcsec of the nucleus.
\item Between 5 and 18\,arcsec from the nucleus the main jet is very bright 
(Figs~\ref{fig:ilo}a -- c).
\item Further out, the brightness of the main jet decreases monotonically, but
the longitudinal intensity profile flattens with increasing distance
(Fig.~\ref{fig:ilo}c).
\item The counter-jet peaks twice, at 15 and 30\,arcsec (Figs~\ref{fig:ilo}a
-- c).
\item The jet/counter-jet sidedness ratio is high within 30\,arcsec of the
  nucleus, thereafter maintaining a constant value  on-axis
  (Fig.~\ref{fig:ilo}).
\item The sidedness ratio is higher on-axis than at the edges of the jets 
(Figs~\ref{fig:ilo}e, f, h, i; Fig.~\ref{fig:transside}).
\end{enumerate}

\subsection{Linear polarization}

\begin{figure*}
\includegraphics[width=17cm]{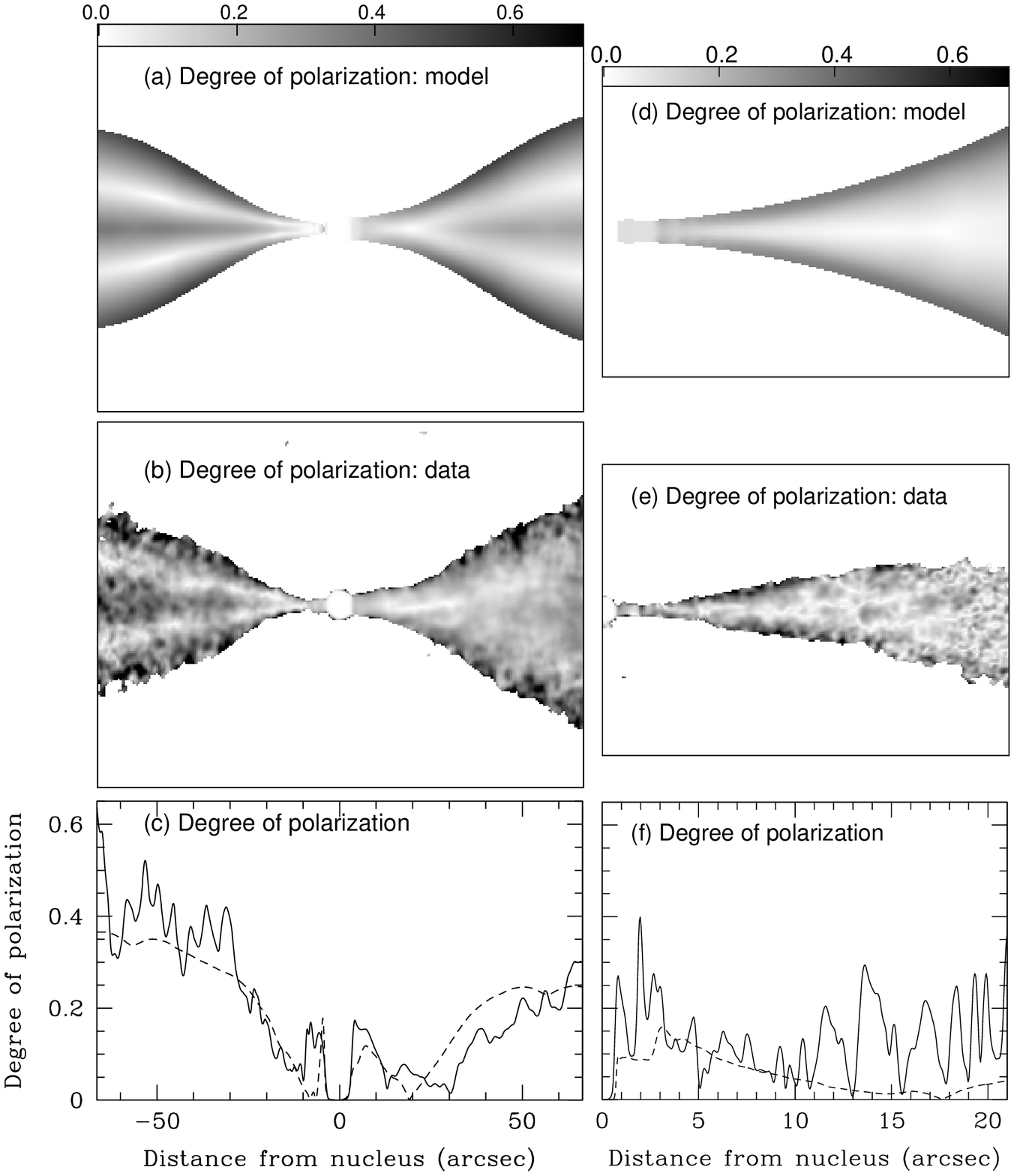}
\caption{A comparison of the degree of polarization, $p$, for model and data.  
(a) and (b) grey scales of $p$ at 2.35\,arcsec resolution in the range 0 -- 0.7.
(c) Profiles of $p$ along the jet axis for the data (full line) and model
  (dashed). Panels (a) -- (c) show the inner 66.5\,arcsec of each jet. (d) and
  (e) grey-scales of $p$ for the main jet at 0.4\,arcsec resolution. (d) model,
  (e) data. (f) Profile of $p$ along the axis for the main jet. Panels (d) --
  (f) show the inner 21\,arcsec of the main jet.   The data are blanked 
  only if $I < 5\sigma_I$, to avoid discriminating against areas of low $p$.
  The models are not blanked.
\label{fig:pol}}
\end{figure*}

Fig.~\ref{fig:pol} compares the model and observed degrees of polarization, $p$
at resolutions of 2.35\,arcsec (a -- c) and 0.4\,arcsec (d -- f). The values of
$p$ are represented as grey-scales and as profiles along the jet axis.
Fig.~\ref{fig:transpol} compares the average transverse profiles of the degree of
polarization at distances from the nucleus between 45 and 66.5\,arcsec. Finally,
Figs~\ref{fig:ivec}, \ref{fig:ivechi} and \ref{fig:ivec.model.small} represent
both the value of $p$ (vector length) and the direction of the apparent magnetic
field (vector direction) at the two resolutions.

The following polarization features are well described by the model:
\begin{enumerate}
\item The degree of polarization in the brightest portion of the main jet has a
  V-shaped structure, with high polarization close to the edges and low
  polarization on-axis (Figs~\ref{fig:pol}d and e).
\item Both jets show a high degree of polarization at their edges, with the
apparent magnetic field parallel to the edge everywhere (Fig.~\ref{fig:ivec}).
\item The main jet shows a transition from longitudinal to transverse apparent
  field on-axis at a distance of $\approx$30\,arcsec from the nucleus whereas
  the counter-jet shows transverse apparent field on-axis everywhere there is
  adequate signal (Fig.~\ref{fig:ivec}).
\item Within $\approx$3\,arcsec of the nucleus, the counter-jet has a transverse
apparent field with $p \approx$ 0.15 -- 0.2. At 3.6\,arcsec, there is a
polarization minimum (consistent with zero); at larger distances the degree of
polarization increases with distance (Fig.~\ref{fig:pol}c). The polarization
direction is difficult to see in Fig.~\ref{fig:ivec}(a), where the model is
blanked if $I < 5\sigma_I$ or $P <3\sigma_P$ to match the data. Close to the
nucleus, the counter-jet has discrete knots of emission with detectable
polarization; the model is smoother, has a lower peak intensity and is totally
blanked.  For this reason Fig.~\ref{fig:ivec.model.small} shows model
polarization vectors close to the nucleus with minimal blanking.
\item The counter-jet shows a much more pronounced ridge of high, transverse
  polarization on-axis than does the main jet (Fig.~\ref{fig:transpol}).
\item The on-axis polarization of the counter-jet is
  systematically higher than that of the main jet at distances $>$30\,arcsec,
  where both show transverse apparent field (Fig.~\ref{fig:pol}c).
\end{enumerate}

\subsection{Features that are not fitted well}
\label{badfit}

As discussed in more detail by Worrall et al.\ (in preparation), the bright
region of the main jet is resolved into complex knots and filaments and its
structure is clearly not axisymmetric in either total intensity or linear
polarization (the same is almost certainly true of the counter-jet, but we see
only its brightest few knots). Our model fits the inner jets with a smooth
brightness distribution of the correct mean value, but there are large local
deviations. An inevitable consequence is that there are fluctuations in
sidedness ratio, which we cannot fit. We emphasize that relying on the sidedness
ratio alone to derive jet properties is dangerous: it is necessary to average
the intensity over a region large enough to contain many fluctuations, but small
compared with variations in the underlying flow. This may not always be
possible.  The intensity fits close to the nucleus have two problems which may
be due to small-scale fluctuations:
\begin{enumerate}
\item There is significant emission from the counter-jet close to the nucleus,
   (Fig.~\ref{fig:ismall}) and the sidedness ratio derived by integrating over
   the emission between 0.8 and 1.5\,arcsec from the nucleus at 0.4\,arcsec
   resolution is 6.65, much smaller than is seen further out 
   (Fig.~\ref{fig:ilo}g).  In principle, such an increase of sidedness cannot
   occur in a monotonically decelerating jet and our model fails to reproduce
   it. We return to this point in Section~\ref{acceleration}.
\item The first brightening of the counter-jet is slightly, but significantly
  further from the nucleus than the corresponding feature in the main jet: the
  model predicts that they should both brighten at the same place
  (Fig.~\ref{fig:ilo}d). This effect causes the huge peak in observed
  sidedness ratio seen in Fig.~\ref{fig:ilo}(g). It almost certainly results
  from the coincidence of a knot in the main jet with a minimum in the
  counter-jet at $\approx$9\,arcsec from the nucleus (Fig.~\ref{fig:ismall}).
\end{enumerate}
A related problem causes a discrepancy between the predicted and observed
degrees of polarization observed on-axis in the bright region of the main jet:
this appears to be due to filaments with aligned apparent fields crossing the
jet axis at an oblique angle (Fig.~\ref{fig:ivechi}b).  At high resolution, the
model fits the lower bound of the polarization profile quite well
(Fig.~\ref{fig:pol}f).

In addition to small-scale, non-axisymmetric structure (which affects all of the
sources we have studied at some level) the model of NGC\,315 also fails to fit
two other features.
\begin{enumerate}
\item The on-axis sidedness ratio is slightly higher than predicted at large
  distances from the nucleus, leading to a larger contrast in sidedness ratio
  between centre and edge than is predicted by the model (Figs~\ref{fig:ilo} and
  \ref{fig:transside}). This may indicate that the transverse velocity profile
  is more complex than we assume, and we discuss this point further in
  Section~\ref{vtrans}. 
\item The average transverse polarization profile for the main jet is slightly
  flatter than predicted by the model at distances between 45 and 66.5\,arcsec
  from the nucleus (Fig.~\ref{fig:transpol}).
\end{enumerate}

\section{Physical parameters}
\label{physical}

\subsection{Summary of parameters}

In this section, all distances in linear units are measured in a plane
containing the jet axis (i.e.\ {\sl not} projected on the sky). The parameters
of the best-fitting model and their approximate uncertainties are given in
Table~\ref{tab:results}.
\begin{table*}
\caption{Fitted parameters and error estimates.\label{tab:results}}
\begin{minipage}{120mm}
\begin{tabular}{llrrr}
\hline
Quantity                 &     Symbol     &~opt~ 
&~min\footnote{The Symbol $<$  means that
 any value smaller than the quoted maximum is allowed.}  
&~max\\
\hline
Angle to line of sight (degrees)& $\theta$& 37.9 & 36.1 & 40.7 \\
&&&&\\
Geometry                                  &&&&\\
~~Boundary position (kpc)       &  $r_0$  &49.18 &47.37 &50.61 \\
~~Jet half-opening angle (degrees)&$\xi_0$         
                                          & 3.75 & 0.08 & 6.52 \\
~~Width of jet at outer boundary (kpc)&$x_0$& 12.71&12.10 &13.37 \\
&&&&\\
Velocity                                  &&&&\\
~~Boundary positions (kpc)                &&&&\\
~~~~inner                &$\rho_{\rm v_1}$&7.59  &4.28  &10.10 \\
~~~~outer                &$\rho_{\rm v_0}$&18.07 &16.24 &19.74 \\
~~On~$-$~~axis velocities / $c$           &&&&\\
~~~~inner                &   $\beta_1$    & 0.88 & 0.77 & 0.99 \\
~~~~outer                &   $\beta_0$    & 0.38 & 0.35 & 0.41 \\
~~Fractional velocity at edge of jet      &&&&\\
~~~~inner                &     $v_1$      & 0.79 & 0.59 & 1.05 \\
~~~~outer                &     $v_0$      & 0.58 & 0.45 & 0.71 \\
&&&&\\
Emissivity                                &&&&\\
~~Boundary positions (kpc)                &&&&\\
~~~~inner                &$\rho_{\rm e_1}$&2.52  & 0.00 & 3.22 \\
~~~~2                    &$\rho_{\rm e_2}$&3.53  & 2.70 & 4.75 \\
~~~~3                    &$\rho_{\rm e_3}$&9.41  & 9.14 & 9.73 \\
~~~~4                    &$\rho_{\rm e_4}$&10.05 & 9.71 & 10.30\\
~~On~$-$~~axis emissivity exponents       &&&&\\
~~~~inner                &     $E_{\rm in}$      & 3.15  & $<$ & 4.00  \\
~~~~2                    &     $E_{\rm rise}$      & 0.03  & $<$ & 2.34  \\
~~~~3                    &     $E_{\rm mid}$      & 2.79  & 2.33  & 3.18  \\
~~~~4                    &     $E_{\rm fall}$      & 10.39 & 6.45  & 13.47 \\
~~~~5                    &     $E_{\rm out}$      & 2.87  & 2.72  &  3.04 \\
~~Fractional emissivity at edge of jet    &&&&\\
~~~~inner boundary       &     $e_1$      & 1.01 & 0.41 & 2.12 \\
~~~~boundary 3          &     $e_0$      & 0.45 & 0.30 & 0.64 \\
&&&&\\
B-field                                   &&&&\\
~~Boundary positions (kpc)                &&&&\\
~~~~inner                &$\rho_{\rm B_1}$&0.46  & 0.00 & 6.47 \\
~~~~outer                &$\rho_{\rm B_0}$&25.79 &20.85 & 31.73\\
~~RMS field ratios                        &&&&\\
~~~~radial/toroidal                       &&&&\\
~~~~~~inner region axis  &     $j_1^{\rm axis}$ &1.12 & 0.54 &   1.71     \\
~~~~~~inner region edge  &     $j_1^{\rm edge}$ &0.45 & 0.03 &   0.82     \\
~~~~~~outer region axis  &     $j_0^{\rm axis}$ &0.61 & 0.08 &   0.91     \\
~~~~~~outer region edge  &     $j_0^{\rm edge}$ &0.20 & 0.00 &   0.39     \\
~~~~longitudinal/toroidal                 &&&&\\                    
~~~~~~inner region axis  &     $k_1^{\rm axis}$ &1.43 & 1.18 &   1.70     \\
~~~~~~inner region edge  &     $k_1^{\rm edge}$ &0.95 & 0.77 &   1.14     \\
~~~~~~outer region axis  &     $k_0^{\rm axis}$ &0.97 & 0.79 &   1.19     \\
~~~~~~outer region edge  &     $k_0^{\rm edge}$ &0.37 & 0.19 &   0.52     \\
\hline						                    
\end{tabular}
\end{minipage}
\end{table*}

\begin{figure}
\includegraphics[width=8.5cm]{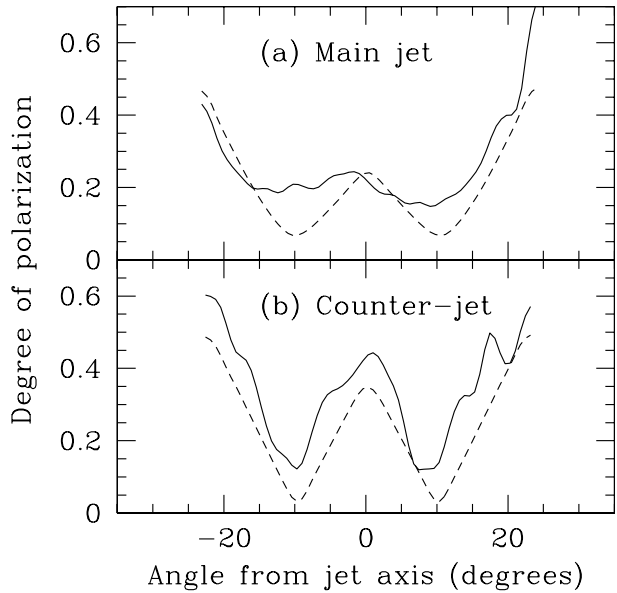}
\caption{Average transverse profiles of the degree of polarization for the main and
  counter-jets at a resolution of 2.35\,arcsec. The profiles are generated by
  averaging along radii between 45 and 66.5\,arcsec from the nucleus and plotting
  against angle from the jet axis. In both panels, the data are represented by
  full lines and the model by dashed lines. (a) main jet, (b) counter-jet.
\label{fig:transpol}}
\end{figure}

\subsection{Geometry and angle to the line of sight}
\label{results:geom}

The best-fitting model has an angle to the line of sight of $\theta = 38^\circ
\pm 2^\circ$, consistent with the range $30^\circ \la \theta \la 40^\circ$
estimated by \citet{Giov01} from the core prominence, proper-motion measurements
and jet/counter-jet intensity on pc scales (the last using an isotropic emission
model). The shape of the outer edge is shown in Fig.~\ref{fig:profiles}(a).  The
initially well-collimated jets flare in the inner 15\,kpc to a maximum opening
angle $\approx$20$^\circ$, which is maintained over most of the rest of the
modelled region. The cubic fit to the jet radius requires that a conical region
should start at a distance of 49\,kpc from the nucleus (90\,arcsec in projection
on the sky) and have an opening angle $<$6$^\circ$. This is consistent with the
observed appearance of the jets at larger distances (Fig.~\ref{fig:modelarea}),
taking into account the bends at 70\,arcsec.  NGC\,315's jets are therefore
similar to those in other objects in recollimating to become almost cylindrical,
but only after the slight bends which limit our modelling. On scales $\ga$500
arcsec in projection, there is a second rapid expansion \citep{Willis81,Brid82}.

\subsection{Velocity}
\label{results:vel}

\subsubsection{Model fits}

The boundaries between the three velocity regimes are plotted, together with
their errors, in Fig.~\ref{fig:profiles}(a).  The on-axis velocity profile,
shown by the full line in Fig.~\ref{fig:profiles}(b), is fit with a constant
value of $\beta = 0.88 \pm 0.1$ over the inner 7\,kpc, but this value is
determined primarily by the data at distances $\ga$2.5\,kpc, where the jet
widens and brightens.  Between 7 and 20\,kpc, the jet decelerates uniformly,
reaching a well-constrained asymptotic speed that is still appreciably
relativistic ($\beta = 0.38 \pm 0.03$) and persists to the end of the modelled
region at 38\,kpc. The best-fitting velocity at the jet edge is slower than the
on-axis value, the edge/on-axis ratio varying from $0.8 \pm 0.2$ close to the
nucleus to $0.58 \pm 0.13$ at large distances. A constant value of 0.6 -- 0.7
would fit quite well everywhere, but a constant (top-hat) transverse profile is
also consistent with the data close to the nucleus.  The full velocity field is
shown as contours in Fig.~\ref{fig:velcont}.

\begin{figure*}
\includegraphics[width=14cm]{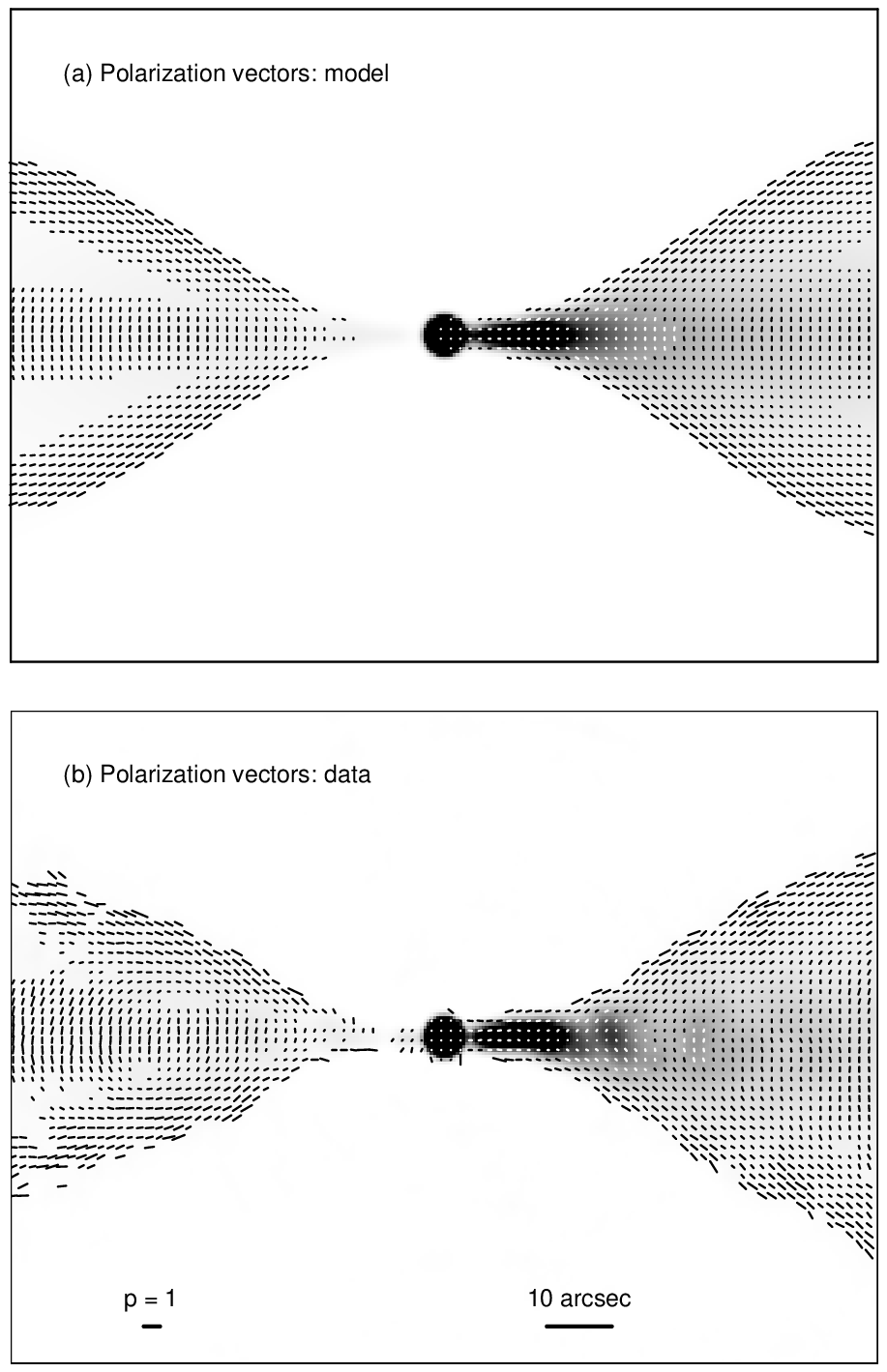}
\caption{Vectors with lengths proportional to the degree of polarization, $p$,
and directions along the apparent magnetic field, superimposed on grey-scales of
total intensity. The resolution is 2.35\,arcsec and vectors are plotted every
1.5\,arcsec. The polarization and angular scales are indicated by the labelled
bars in the lower panels and the areas plotted are the same as those in
Figs~\ref{fig:ilo} and \ref{fig:pol}. Vectors are plotted only where $I >
5\sigma_I$ and $P > 3\sigma_P$ (Table~\ref{noise}). (a) model; (b) data.
\label{fig:ivec}}
\end{figure*}

\begin{figure*}
\includegraphics[width=10cm]{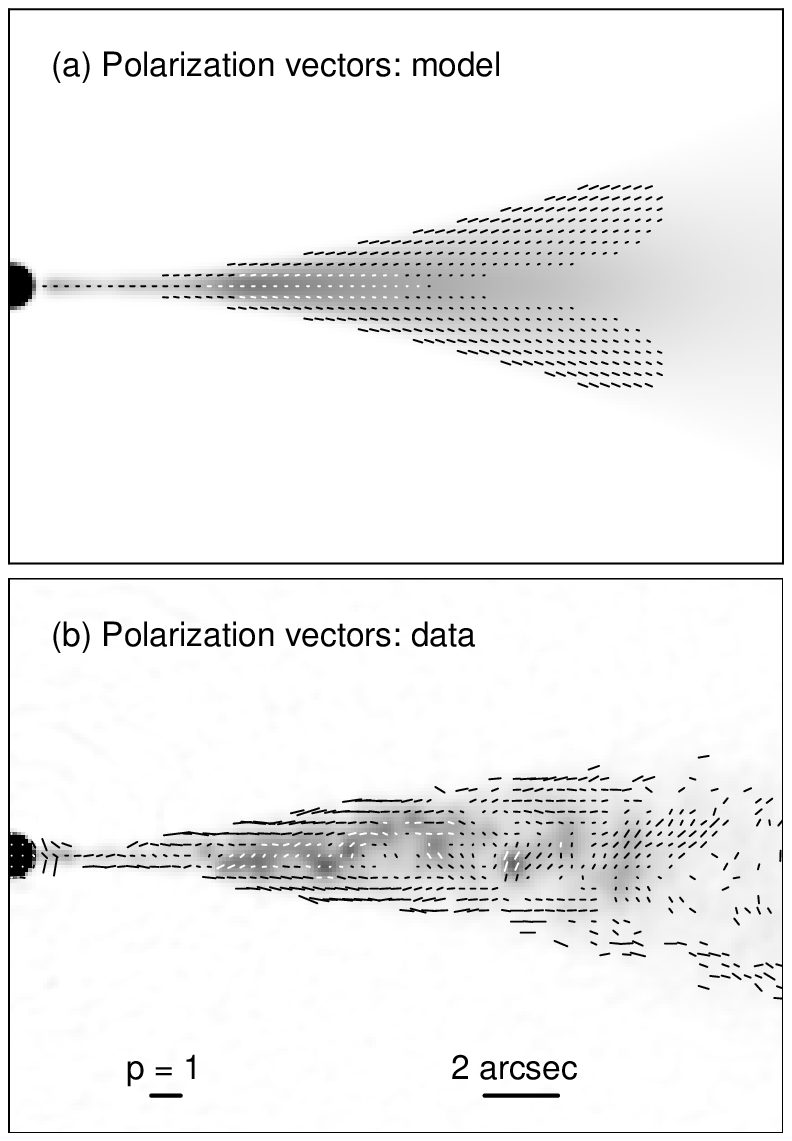}
\caption{Vectors with lengths proportional to the degree of polarization, $p$,
and directions along the apparent magnetic field, superimposed on grey-scales of
total intensity. The resolution is 0.4\,arcsec FWHM and vectors are plotted
every 0.4\,arcsec. The polarization and angular scales are indicated by the
labelled bars in the lower panels and the areas plotted are the same as those in
Figs~\ref{fig:ihi} and \ref{fig:pol}.  Vectors are plotted only where $I >
5\sigma_I$ and $P > 3\sigma_P$ (Table~\ref{noise}). (a) model; (b) data.
\label{fig:ivechi}}
\end{figure*}

\subsubsection{Evidence for acceleration?}
\label{acceleration}

On pc scales, both the jet/counter-jet ratio and the apparent component speed
increase with distance from the nucleus \citep{Cotton99,Giov01}, although no
changes in velocity have yet been detected in any individual component. The
velocities derived from the measurements in \citet{Cotton99}, but taking our
best-fitting inclination angle of $\theta = 37.9^\circ$ and $H_0$ =
70\,$\rm{km\,s^{-1}\,Mpc^{-1}}$, are plotted in Fig.~\ref{fig:vcomp}. The
velocities from $I_{\rm j}/I_{\rm cj}$ are systematically higher than those from
component motions at the same distance from the nucleus for this Hubble
Constant.  An estimate from $I_{\rm j}/I_{\rm cj}$ for our 0.4\,arcsec image
between 0.8 and 1.5\,arcsec from the core in projection ($\approx$0.6\,kpc along
the axis) is also included. This assumes (as in \citealt{Cotton99}) that the
rest-frame emission is isotropic, so that:
\[
\frac{I_{\rm j}}{I_{\rm cj}} = \left(
\frac{1+\beta\cos\theta}{1-\beta\cos\theta} \right) ^{2+\alpha}
\]
We also plot the results of our model fits for the centre and edge of the jet
for distances $>$2.5\,kpc, where they are well determined. Aside from the low
value at $\approx$0.6\,kpc, the velocities are consistent with acceleration from
$\beta \approx 0.7$ at 1\,pc to $\beta \approx 0.95$ at 10\,pc, an approximately
constant velocity between 10\,pc and 8\,kpc and deceleration as described
earlier.  M\,87 \citep{Reid,Biretta95,BJ95,JB95} and Cen\,A \citep{TPJ01,Hard03}
also show component speeds which increase from pc to kpc scales. Similarly,
NGC\,6251 has been suggested as an example of an accelerating flow because its
jet/counter-jet ratio increases with distance \citep{Sudou}, although the
detection of a counter-jet on pc scales was not confirmed by \citet{JW}. Two
possible mechanisms for increase of velocity on pc scales are thermal
acceleration of proton-electron plasma \citep{MLF} and magnetic driving
\citep{VK}.

As noted by \citet{Cotton99}, it is unclear whether the apparent acceleration on
pc scales occurs because the bulk flow accelerates or because we see different
parts of a jet stratified in velocity at different distance from the core.  An
acceleration process which only reaches a speed of $\beta \approx 0.7$ at 1\,pc
is impossible to reconcile with FR\,I radio galaxies being the parent population
of TeV blazars, as highly relativistic ($\Gamma \ga$ 10 -- 20) flow is required
on 0.1-pc scales in the latter class to produce the observed $\gamma$-ray
emission. Velocity gradients have indeed been inferred on sub-pc scales to
explain the simultaneous observations of slowly-moving radio components and
highly variable TeV emission in blazars \citep{Ghis05}. Similarly, although the
anomalously small velocity estimate at 0.6\,kpc in NGC\,315 might well result
from a random intensity fluctuation, 3C\,31 (LB) and B2\,0326+39 (CL) also show
lower sidedness ratios close to the nucleus than at the flaring point,
suggesting that the material contributing the bulk of the emission in the faint
region at the jet base really is slower than that further out, where our models
are well constrained.  We cannot resolve the transverse velocity structure of
the jets in their innermost regions, so it is possible that they have very fast
central spines whose emission is Doppler dimmed on both sides of the nucleus and
that the visible emission comes from much slower surface layers. Rapid
deceleration of the spine to a speed $\beta_1 \approx 0.9$ at 2.5 -- 3.5\,kpc
from the nucleus would make it visible, dominating the emission and causing the
sidedness ratio to rise, as suggested for 3C\,31 by LB.
Further from the nucleus, there are local minima in the sidedness ratio at 35
and 50\,arcsec, separated by a maximum at 40\,arcsec (Fig.~\ref{fig:ilo}j). These
could be interpreted as changes in velocity in the range $0.32 < \beta_0 <
0.42$, but they show no coherent pattern, so we believe that they probably
reflect intrinsic differences between the two jets.

\begin{figure}
\includegraphics[width=8.5cm]{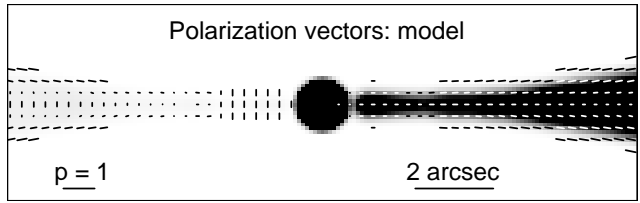}
\caption{Vectors with lengths proportional to the degree of polarization, $p$,
and directions along the apparent magnetic field, superimposed on grey-scales of
total intensity for the model at a resolution of 0.4\,arcsec FWHM.  The
polarization and angular scales are indicated by the labelled bars and the inner
$\pm$8\,arcsec is shown.  Vectors are plotted if $I > 2$\,$\mu$Jy\,(beam
area)$^{-1}$ (cf.\ Fig.~\ref{fig:ivechi}). This plot shows the model
polarization structure of the counter-jet near the nucleus, including the null
at 3.6\,arcsec and the region of transverse apparent field closer in. The data
are not plotted for comparison at this resolution because the signal-to-noise
ratio in the counter-jet is too low, but these features are qualitatively
consistent with the low-resolution images (Figs~\ref{fig:pol}b,c and
~\ref{fig:ivec}b). 
\label{fig:ivec.model.small}}
\end{figure}

\subsubsection{Transverse velocity profile}
\label{vtrans}

The discrepancy between the predicted and observed transverse profiles of
jet/counter-jet sidedness ratio (Fig.~\ref{fig:transside}) provides the first
evidence that the Gaussian and spine/shear-layer velocity profiles we have
employed so far may be oversimplified. The effect, which is present in all three
average profiles shown in Fig.~\ref{fig:transside}, is that the sidedness ratio
has a sharp peak within $\approx 5^\circ$ of the jet axis, drops abruptly at
$\approx 10^\circ$ and has a flatter wing at the edge of the jet.  The type of
velocity profile which could produce the observed results would have a central
spine of high velocity ($\beta \approx 0.5$) surrounded in turn by a relatively
narrow shear layer and an outer wing with $\beta \approx 0.2$. There are
important implications for the evolution of the magnetic field, which would be
sheared only over a narrow range of radii, rather than the majority of the jet,
and for the physics of the deceleration process.  The other sources we have
studied (LB, CL) show no systematic discrepancies, but are less well resolved
than NGC\,315 in regions where they have significant changes in sidedness from
axis to edge.

\subsection{Emissivity}
\label{results:emiss}

\begin{figure*}
\includegraphics[width=17cm]{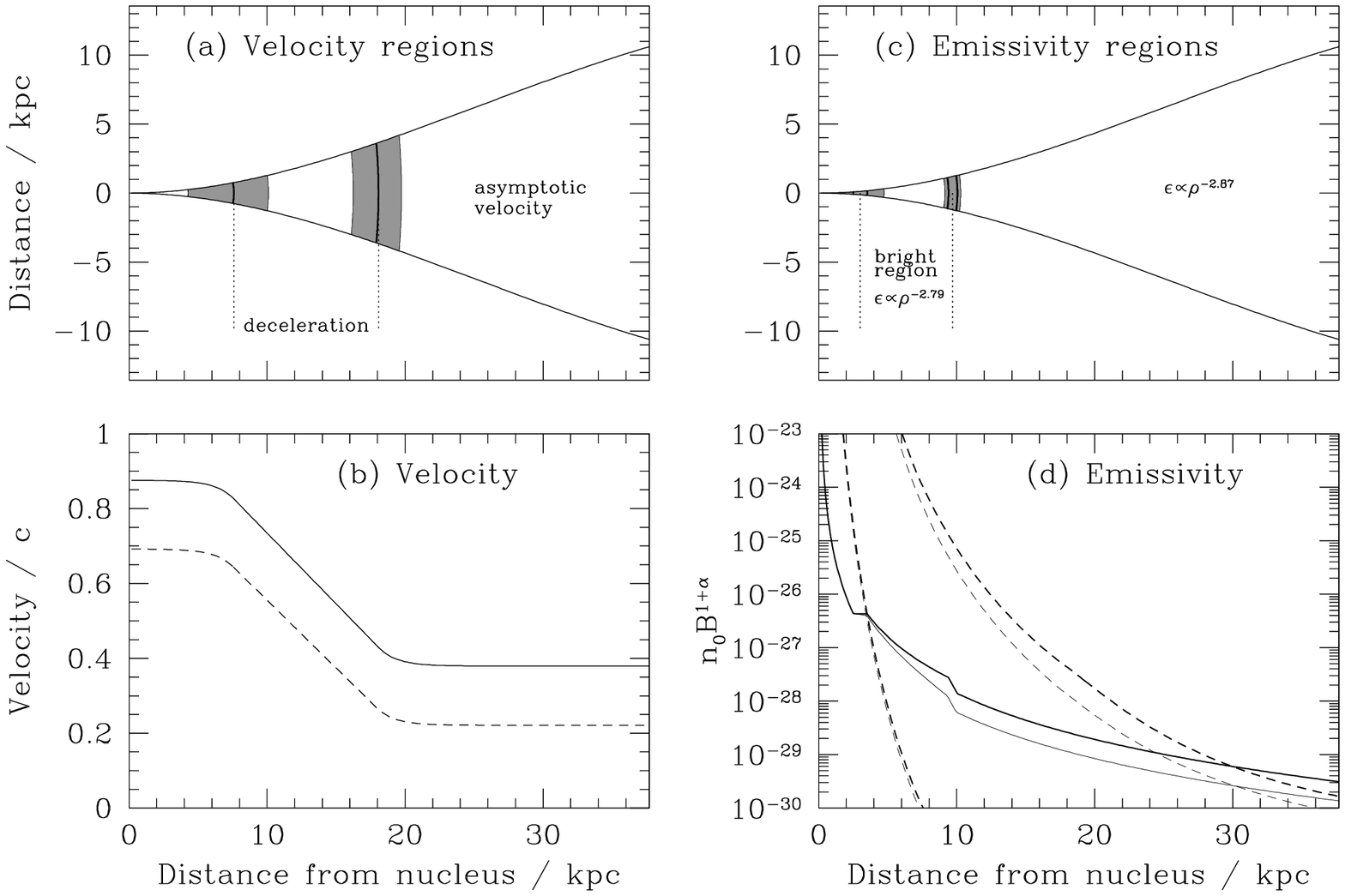}
\caption{Panels (a) and (c): sketches showing the relative positions of the
boundary surfaces between velocity and emissivity regions. The boundaries are
defined in Table~\ref{tab:param} and their positions for the best fitting model
are given in Table~\ref{tab:results}. The full vertical curves indicate the
boundaries and the shaded areas their allowed ranges, also from
Table~\ref{tab:results}. (a) Velocity. The regions of approximately uniform
deceleration and asymptotic outer velocity are marked. (c) Emissivity. The
region of enhanced emissivity between 3 and 10\,kpc is indicated.  (b) and (d):
profiles of intrinsic parameters along the jets in the rest frame. (b) the
velocity profile along the jet axis (solid line) and jet edge (dashed line). (d)
$n_0 B^{1+\alpha}$ derived from the emissivity, with $n_0$ and $B$ in SI
units. Solid line: model; dashed line, adiabatic approximation with the
magnetic-field structure expected from flux freezing. The adiabatic model curves
are plotted twice, normalized to match the model at 3.5 and 30\,kpc from the
nucleus, respectively. The heavy lines show the on-axis profiles; thin lines the
edge profiles.
\label{fig:profiles}}
\end{figure*}

\begin{figure}
\includegraphics[width=8.2cm]{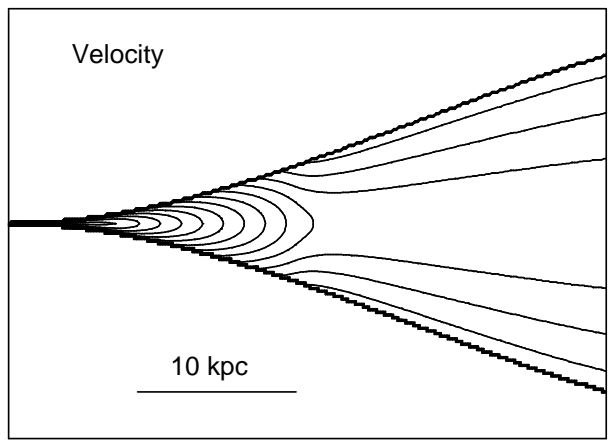}
\caption{Contours of the model velocity field. The contours are at intervals of
  0.05 in the range $\beta =$ 0.25 -- 0.85.
\label{fig:velcont}}
\end{figure}

\begin{figure}
\includegraphics[width=8.5cm]{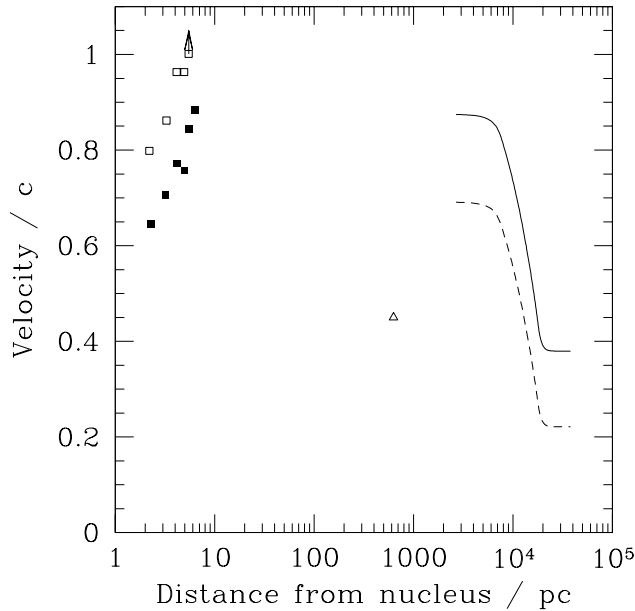}
\caption{A comparison of velocity estimates on pc and kpc scales, plotted
  against distance from the nucleus (note the logarithmic scale). We assume that
  $\theta = 37.9^\circ$ everywhere. Filled squares: velocities from proper
  motions; open squares: velocities from jet/counter-jet ratios (both from
  \citealt{Cotton99}, but with our choice of Hubble Constant). The open 
  triangle shows the velocity derived from the jet/counter-jet ratio at
  0.4\,arcsec resolution close to the nucleus. All velocity estimates from
  intensity ratios alone are calculated for isotropic emission in the rest
  frame. The full and dotted lines show our model fits for the centre and edge
  of the jets, respectively. They are plotted only for distances $>$2.5\,kpc,
  where they are well constrained (see Section~\ref{acceleration}).
\label{fig:vcomp}}
\end{figure}

The boundaries between the emissivity regions, with their errors, are shown in
Fig.~\ref{fig:profiles}(c) and the on-axis profile of $n_0 B^{1+\alpha}$,
derived from the emissivity, is plotted as the heavy, full line in
Fig.~\ref{fig:profiles}(d). $n_0$ and $B$ are in SI units.  The emissivity
profile at distances $\la$3.5\,kpc from the nucleus is poorly constrained
(Table~\ref{tab:results}). Again, this is partly because the jets are faint and
poorly resolved, but more as a consequence of the poor fit to the counter-jet
(Sections~\ref{badfit}), which causes the $\chi^2$ values to change very little
as the model parameters are varied. The profile plotted in
Fig.~\ref{fig:profiles}(d) is essentially determined by the brightness
distribution of the main jet for distances $<$3.5\,kpc. We model it as an inner
region with a slope $E_{\rm in} \approx 3$ ($<$2.5\,kpc) followed by a
transition zone with a much flatter slope $E_{\rm rise} \approx 0$ from 2.5 --
3.5\,kpc. The sudden increase in the observed surface brightness of the main jet
(Fig.~\ref{fig:ihi}) is then produced by a rapid expansion of the jet at roughly
constant velocity and emissivity.  The emissivity profile for the bright,
well-resolved sections of the jets can be divided into two power-law sections,
with indices of $E_{\rm mid} = 2.8$ (3.5 -- 9.4 kpc) and $E_{\rm out} = 2.9$
($>$10.1\,kpc). These are separated by a second short transition zone over which
the emissivity drops by a factor $\approx$2, modelled as a very steep power law
with index $E_{\rm fall} \approx 10.4$ between 9.4 and 10.1\,kpc; this might
also be represented as a discontinuity. The {\em bright region} (3.5 -- 9.4 kpc)
contains complex, non-axisymmetric and knotty structure (Figs~\ref{fig:ihi}b and
\ref{fig:ivechi}b), whose average is represented well but whose details are not.
The outer section is, by contrast, relatively smooth. There is no evidence for
any transverse variation of emissivity at the start of the bright region,
although this is poorly constrained. In the outer section the edge emissivity is
about half of its on-axis value (a profile of $n_0 B^{1+\alpha}$ at the jet edge
is shown as the light, full line in Fig.~\ref{fig:profiles}c).

The bright region is clearly differentiated from the rest because its emissivity
is a factor of 2 higher than expected from a smooth extrapolation between
smaller and larger distances (Fig.~\ref{fig:profiles}d). We also resolve the
inner boundary of this region in 3C\,31 and B2\,0326+39 (LB, CL), modelling it
as a discontinuous increase in emissivity.  The bright region comes
to an equally abrupt end, marked by an almost discontinuous drop in emissivity
in NGC\,315 (Fig.~\ref{fig:profiles}d) and B2\,0326+39 (fig.~16c of
CL). B2\,1553+24 may show a similar feature, but is less well resolved (fig.~18c
of CL). In all three sources, the drop in emissivity occurs well before the jets
recollimate. 3C\,31 does not show any sharp decrease in emissivity.

The phenomenon of sudden brightening and expansion at a {\em flaring point}
close to the nucleus is very common in FR\,I jets (e.g. \citealt{Parma87}), but
the equally sudden emissivity drop at the end of the bright region has only
become apparent from our modelling of NGC\,315 and B2\,0326+39. In both sources,
the drop is located just after the start of the rapid deceleration ($\rho =
\rho_{v_1}$; see Table~\ref{tab:results}) and coincides with it to within the
errors. This association reinforces the argument that dissipation in FR\,I jets,
leading to enhanced radio emission and the production of synchrotron radiation
at much higher frequencies, is associated mainly with their fastest parts
\citep{LB04}. 

\subsection{Magnetic-field structure}
\label{field-structure}

\begin{figure*}
\includegraphics[width=14cm]{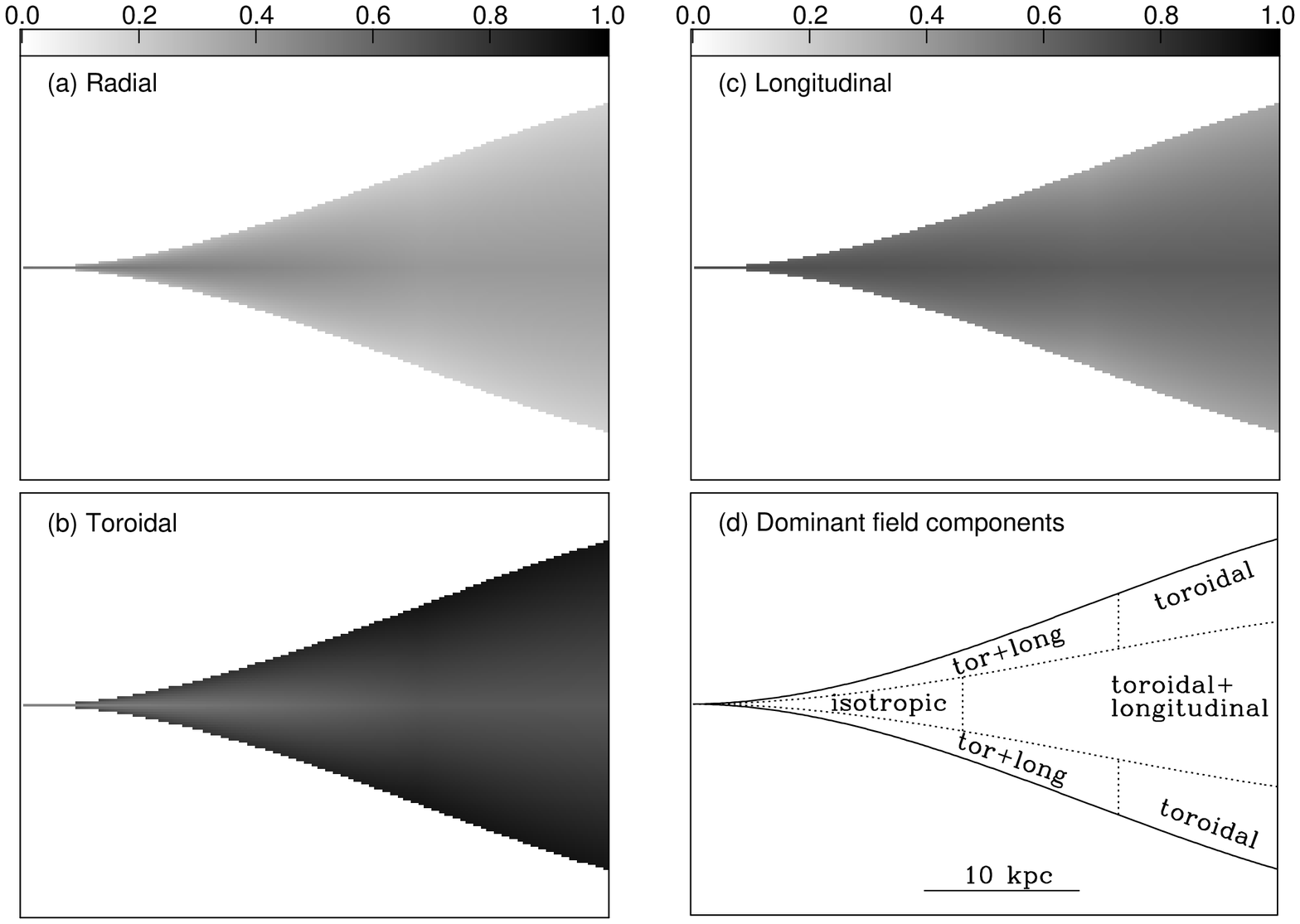}
\caption{Panels (a) -- (c): grey-scales of the fractional magnetic-field
  components. (a) radial, $\langle B_r^2 \rangle^{1/2}/B$; (b) toroidal $\langle
  B_t^2 \rangle^{1/2}/B$; (c) longitudinal $\langle B_l^2 \rangle^{1/2}/B$. $B =
  \langle B_r^2+B_t^2+B_l^2 \rangle^{1/2}$. (d) Sketch of the dominant field
  components at different locations in the jet, as deduced from the model fits.
\label{fig:bgrey}}
\end{figure*}

The variation of magnetic-field structure is illustrated by
Fig.~\ref{fig:bgrey}, which shows grey-scales of the fractional field
components: radial, $\langle B_r^2 \rangle^{1/2}/B$, toroidal, $\langle B_t^2
\rangle^{1/2}/B$ and longitudinal, $\langle B_l^2 \rangle^{1/2}/B$ as defined in
Section~\ref{field-parms}. Profiles of these quantities as functions of distance
along the jet axis, $z$, are plotted in Fig.~\ref{fig:bprofiles} for on-axis and
edge streamlines.  All three components vary slowly out to a distance of
25\,kpc, after which they remain constant. The errors on the radial component
are much larger than on the other two. On-axis, all three components are
initially roughly equal (i.e. the field is on average isotropic); further out
the toroidal component increases and the radial and longitudinal 
components decrease. At the edge of the jet, the toroidal component is always
the largest, and dominates over the other two at large distances
(Fig.~\ref{fig:bprofiles}d).  The main field components in different parts
of the jets are shown schematically in Fig.~\ref{fig:bgrey}(d).

It is difficult to assess the uncertainties in the field component ratios from
the error estimates in Table~\ref{tab:results}, so each panel of
Fig.~\ref{fig:bprofiles} includes a shaded area defining the region which the
profile could occupy if any one of the six free parameters defining it is varied
up to its quoted error.

The field structure we infer for the jets in NGC\,315 within $\approx$15\,kpc of
the nucleus shows some similarities to that suggested for M\,87 by
\citet{Perlman}. In our model for NGC\,315, the best estimate for the field
configuration on-axis in this region is roughly isotropic (although there are
large uncertainties in the magnitude of the radial component;
Fig.~\ref{fig:bprofiles}b); at any rate, the radial component is larger on-axis
than at the edge, where longitudinal and toroidal components dominate.
\citet{Perlman} suggest that the perpendicular component of {\sl apparent} field
in M\,87 is larger close to the axis and is associated preferentially with the
optical knots whereas the (mainly radio) edge emission has its apparent field
aligned with the jet axis. In our picture, the aligned apparent field at the
edge comes from the projection of intrinsically longitudinal and toroidal
components and the radial component is significant only on-axis.

Finally, we note that symmetry of the average transverse intensity and
polarization profiles rules out a global helical field unless the jet is
observed side-on in the rest frame of the emitting material \citep{Laing81}. The
condition for side-on emission in the rest frame for the approaching jet in
NGC\,315, $\beta = \cos\theta = 0.79$, is roughly satisfied before the jets
decelerate, but not for distances $\ga$10\,kpc from the nucleus. In the
counter-jet, the condition can never hold.  The intensity and polarization
profiles at large distances (Figs~\ref{fig:transi} and ~\ref{fig:transpol}) are
extremely symmetrical, particularly in the counter-jet, and we can rule out
globally-ordered helical fields on these scales.  A configuration in which the
toroidal component is vector-ordered but the longitudinal component has many
reversals would give identical brightness and polarization distributions to
those we calculate for a fully disordered, anisotropic field and we cannot rule
out such a field configuration.

\subsection{Flux freezing and adiabatic models}
\label{ffad}

\begin{figure*}
\includegraphics[width=14cm]{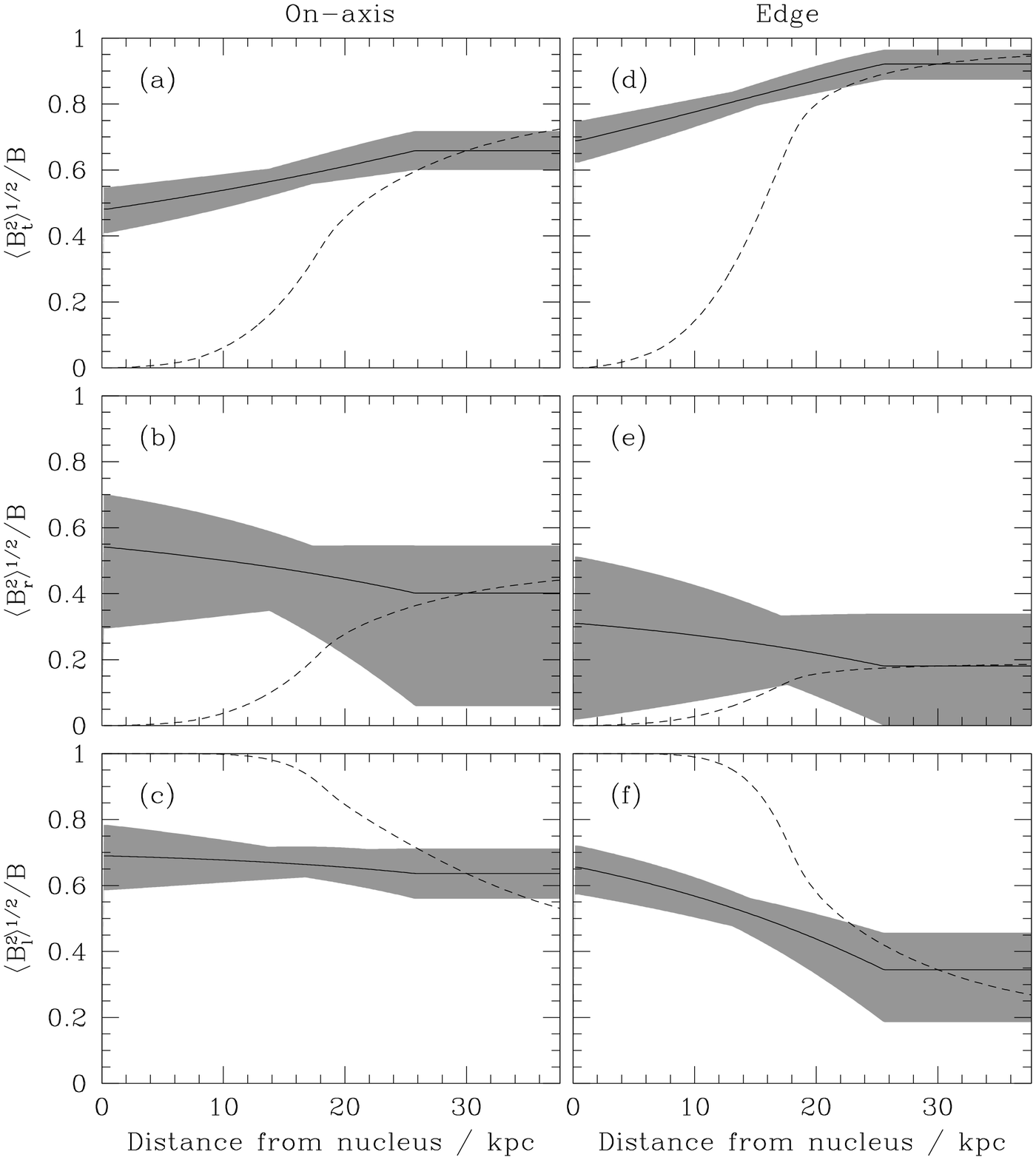}
\caption{Profiles of magnetic-field components along the axis of the jet (panels
  a -- c) and its edge (panels d -- f). The solid lines show the best-fitting
  model, the shaded areas the uncertainties derived from the limits in
  Table~\ref{tab:results} and the dashed lines the profiles expected for a magnetic
  field frozen into the flow.  The profiles for a passively convected field are
  normalized to the free model predictions at a distance of 30\,kpc from the
  nucleus. (a) and (d) toroidal; (b) and (e) radial; (c) and (f) longitudinal.
\label{fig:bprofiles}}
\end{figure*}

In this section, we follow \citet{LB04} in referring to our detailed fits as
{\em free models} in order to distinguish them from the {\em adiabatic models}
considered here.  Given the assumption of flux freezing in a jet without a
transverse velocity gradient, the magnetic field components evolve according to:
\begin{eqnarray*}
B_r &\propto& (x\beta\Gamma)^{-1} \\
B_t &\propto& (x\beta\Gamma)^{-1} \\
B_l &\propto& x^{-2}              \\
\end{eqnarray*}
in the quasi-one-dimensional approximation, where $x$ is the radius of the jet
\citep{Baum97}. The dashed lines in Fig.~\ref{fig:bprofiles} show the predicted
evolution of the field components, normalized to match the models at a distance
of 30\,kpc from the nucleus. The evolution of the longitudinal and toroidal
components is qualitatively as expected but quantitatively inconsistent: the
longitudinal/toroidal ratio decreases with distance, but much less rapidly than
predicted.  Shear will tend to slow the decline of the longitudinal component,
however, so an axisymmetric adiabatic model of the type described by
\citet{LB04} may provide a better description. If the transverse velocity
profile indeed has the form suggested by the average sidedness profile
(Fig.~\ref{fig:transside}, Section~\ref{vtrans}), then shear would be localised
at intermediate radii, a more complex situation than that considered by
\citet{LB04}.  In contrast, the evolution of the on-axis radial component, is
qualitatively inconsistent with flux freezing in {\em any} simple axisymmetric,
laminar-flow model: the radial/toroidal field ratio decreases with distance
instead of remaining constant. Anomalous behaviour of the radial component also
occurs in 3C\,31 and B2\,0326+29 (LB, CL).

Assuming that the radiating electrons suffer only adiabatic losses, and again
adopting the quasi-one-dimensional approximation, the emissivity is:
\begin{eqnarray*}
\epsilon \propto (x^2\beta\Gamma)^{-(1+2\alpha/3)}B^{1+\alpha}
\end{eqnarray*}
\citep{Baum97,LB04}.  $B$ can be expressed in terms of the
parallel-field fraction $f = \langle B_l^2 \rangle ^{1/2}/B$ and the
radius $\bar{x}$, velocity $\bar{\beta}$ and Lorentz factor
$\bar{\Gamma}$ at some starting location using equation 8 of \citet{LB04}:
\begin{eqnarray*}
B \propto \left[ f^2\left(\frac{\bar{x}}{x}\right)^4
+ (1 - f^2)\left(\frac{\bar{\Gamma}\bar{\beta}\bar{x}}{\Gamma\beta
x}\right)^2\right ]^{1/2}
\end{eqnarray*}
We compare the adiabatic and free model emissivity profiles, normalized at 
distances of 3.5 and 30\,kpc, in Fig.~\ref{fig:profiles}(d). In the innermost region,
where the fit to the counter-jet is poor, the slope $E_{\rm in}$ of the
emissivity variation is essentially unconstrained (Section~\ref{results:emiss}
and Table~\ref{tab:results}). Everywhere else there is a clear difference: the
emissivity predicted by the adiabatic approximation falls more rapidly than that
derived from the free model. Dissipative processes must occur to accelerate the
radiating particles where X-ray synchrotron emission is detected, i.e.\ at
distances up to at least 15\,arcsec in projection, corresponding to 8\,kpc in
the jet frame \citep{WBH,DSG}.  The failure of adiabatic models is therefore
inevitable at these distances, but our results suggest that they are inadequate
to describe the emissivity variations anywhere in the modelled region. The
discrepancies are probably too large to be accounted for by field amplification
due to shear in a laminar, axisymmetric, adiabatic model of the type developed
by \citet{LB04}. In general, we have found that adiabatic models only come close
to fitting the brightness distributions of FR\,I jets after they have
recollimated (LB, CL, \citealt{LB04}) and we have not modelled this region in
NGC\,315.

%------------------------- ****** SUMMARY/FURTHER WORK ****** --------------------%

\section{Summary and further work}
\label{ssfw}

\subsection{Summary}
\label{summary}

We have shown that the synchrotron emission from the flaring region of the jets
in the FR\,I radio galaxy NGC\,315 can be fit accurately on the assumption that
they are intrinsically symmetrical, axisymmetric, relativistic decelerating
flows. The functional forms we use to describe the geometry, emissivity,
velocity and magnetic-field structure are very close to those developed in our
previous work (LB, CL). The geometry and the relative locations of the
emissivity and velocity regions are very similar to those in two of
the other sources we have modelled (3C\,31 and B2\,0326+39), except that all of
the physical scales in NGC\,315 are larger by a factor of $\approx$5.

\subsubsection{Geometry}

We have modelled only the {\em flaring region} within 70\,arcsec (in projection)
of the nucleus, as the jets bend shortly thereafter. As in other objects we have
studied, the radius $x$ of its outer isophote is well fitted by the expression
$x = a_2 z^2 + a_3 z^3$, where $z$ is the distance from the nucleus along the
axis.  The jets make an angle of $\theta = 38^\circ \pm 2^\circ$ with the line
of sight, so the size of the region we model is 38\,kpc and the intrinsic length
of the flaring region is $\approx$50\,kpc, much larger than in the other
objects.

\subsubsection{Velocity}

The velocity is well constrained from 2.5\,kpc outwards, where the jet brightens
rapidly (the {\sl flaring point}).  From 2.5 to $\approx$8\,kpc the on-axis
speed is consistent with a constant value of $\beta = 0.88 \pm 0.11$.  This is
very similar to the values derived at the flaring point for the other sources we
have modelled in detail (LB, CL) and from a statistical analysis of sidedness
ratios for a larger sample \citep{LPdRF}.  An anomalous region of low
jet/counter-jet sidedness ratio $<$2.5\,kpc from the nucleus appears to indicate
a lower velocity there, but the jets are faint and poorly resolved, so this
could be due to local fluctuations in the jet or counter-jet brightness. Between
8 and 18\,kpc the jet decelerates uniformly to an asymptotic speed of $\beta =
0.38 \pm 0.03$ which is maintained until the end of the modelled region.
B2\,0326+39 and B2\,1553+24 show velocity profiles of identical form, but with
lower asymptotic velocities (CL), whereas 3C\,31 continues to decelerate slowly
on larger scales (LB). NGC\,315 shows a significant transverse velocity
gradient, with an edge velocity consistent with 0.6 -- 0.7 of the on-axis value
everywhere, as in the other sources. There are hints from the 
sidedness ratio at large distances that our assumed (Gaussian) form for the
transverse velocity profile may be inadequate, and that a profile with a central
spine ($\beta \approx 0.5$) separated from an outer sheath ($\beta \approx 0.2$)
by a relatively narrow shear layer may provide a better fit.

\subsubsection{Emissivity}

The emissivity profile along the jets is modelled as three main power-law
sections with slopes of $-3.5$ (0 -- 2.5\,kpc; very poorly constrained), $-2.8
\pm 0.5$ (3.5 -- 9.4\,kpc) and $-2.9 \pm 0.2$ (9.4 -- 38\,kpc). These are
separated by short transition zones, also modelled as power laws. The first of
these (2.5 -- 3.5\,kpc) is roughly constant and represents the brightening of
the jet as a very rapid expansion at constant emissivity. The second, from 9.4
-- 10.1\,kpc is very steep and describes an almost discontinuous drop in
emissivity by a factor of 2. Simple adiabatic models predict too steep an
emissivity decline, as we also found for the flaring regions of other jets.
The emissivity is centre-brightened where its transverse variation is well
constrained.

\subsubsection{Magnetic field}

To fit the polarization structure of the jets, transverse variation of field
structure had to be included in our models. On axis, the field is roughly
isotropic $\la$10\,kpc from the nucleus, but the radial component declines,
leaving an equal mix of longitudinal and toroidal field by the end of the
modelled region. At the edge of the jets, the radial component is small and the
field configuration evolves from an equal mix of longitudinal and toroidal close
to the nucleus to almost pure toroidal at large distances. From the symmetry
of the transverse intensity and polarization profiles, particularly in the outer
parts of the jets, we infer that there cannot be a significant, globally-ordered
helical field.  All three components could have many reversals or the toroidal
component could be globally ordered, provided that the other two are not.  The
evolution of the radial field component along the jets is not consistent with
flux freezing in our assumed velocity field.  That of the toroidal and
longitudinal components is qualitatively as expected, but a more complex model,
including shear, is required for a quantitative test.

\subsection{Further work}
\label{further work}

We are currently acquiring VLA data for one further source, 3C\,296
\citep{Hardcastle97}. We will then present model fits for all five sources using
the same functional forms in order to compare their properties
quantitatively. We will also investigate more complex transverse velocity
profiles, as outlined in Section~\ref{vtrans} and develop techniques to deal
with slightly bent jets. Where the quasi-one-dimensional analysis presented here
indicates that the adiabatic approximation is reasonable, we will fit the
brightness distributions using the self-consistent adiabatic model of
\citet{LB04}.  Once suitable X-ray observations have been made, we plan to apply
the conservation-law approach of \citet{LB02b} to derive the energy and momentum
fluxes of the modelled jets and their variations of pressure, density and
entrainment rate with distance from the nucleus.

There are, as yet, no observations of jets in FR\,II sources, or in any class of
source on scales $\la$1\,kpc, with resolution and sensitivity adequate for detailed
modelling. The advent of EVLA, e-MERLIN and broad-band VLBI  should
allow us to apply our techniques to more powerful (and probably faster) jets and
to probe scales much closer to those on which jets are launched.

%--------------------------- ****** THE END ****** ------------------------------%
\section*{Acknowledgments}

JRC acknowledges a research studentship from the UK Particle Physics
and Astronomy Research Council (PPARC). The National Radio Astronomy
Observatory is a facility of the National Science Foundation operated
under cooperative agreement by Associated Universities, Inc. We thank the
referee, Paddy Leahy, for a careful reading of the paper.

\label{lastpage}
\end{document}